# Nucleation in confinement generates long-range repulsion between rough calcite surfaces

Joanna Dziadkowiec*,[1], Bahareh Zareeipolgardani[2], Dag Kristian Dysthe[1], Anja Røyne[1]

[1]Physics of Geological Processes (PGP), The NJORD Centre, Department of Physics, University of Oslo, Oslo 0371, Norway

[2]Institut Lumière Matière, Université de Lyon, Université Claude Bernard Lyon 1, CNRS UMR 5586, Campus de la Doua, F-69622 Villeurbanne cedex, France

* joanna.dziadkowiec@fys.uio.no

# ABSTRACT

Fluid-induced alteration of rocks and mineral-based materials often starts at confined mineral interfaces where nm-thick water films can persist even at high overburden pressures and at low vapor pressures. These films enable transport of reactants and affect forces acting between mineral surfaces. However, the feedback between the surface forces and reactivity of confined solids is not fully understood. We used the surface forces apparatus (SFA) to follow surface reactivity in confinement and measure nm-range forces between two rough calcite surfaces in NaCl, $CaCl_2$, or $MgCl_2$ solutions with ionic strength of 0.01, 0.1 or 1 M. We observed long-range repulsion that could not be explained by changes in calcite surface roughness, surface damage, or by electrostatic or hydration repulsion, but was correlated with precipitation events which started at µm-thick separations. We observed a poorly crystalline or amorphous precipitate that formed in the confined solution. This liquid-like precipitate did not undergo any spontaneous ripening into larger crystals, which suggested that confinement prevented its dehydration. Nucleation was significantly postponed in the presence of $Mg^{2+}$. The long-range repulsion generated by nucleation between confined mineral surfaces can have a crucial influence on evolution of the microstructure and therefore the macroscopic strength of rocks and materials.

# Introduction

Fluid-driven mineral reactions in nm- to μm-wide confined spaces can significantly differ from bulk processes as small fluid volumes, slow diffusion and limited advection may promote mineral growth[1]. Reactive mineral contacts at grain boundaries and fracture tips frequently govern the macroscopic mechanical strength of rocks and building materials[2,3]. However, it is not clear what is the relative importance of crystallization and interfacial forces in determining the strength of individual solid-solid contacts. In geological environments, nm-range surface forces are relevant down to several km depth in the subsurface. In these regions, MPa-range positive disjoining pressures[4] (or repulsive forces) can sustain the overburden pressure, and thus nm-thin water films can be maintained between contacting mineral surfaces[5]. Recent experimental[6,7] and modelling studies[8,9] of confined single crystal precipitation suggest that there is a strong link between confined mineral growth and the presence of repulsive surface forces that control the thickness of the water films separating the surfaces. The feedback between surface forces and confined mineral growth needs to be further examined.

Calcite is a major mineral resource and biomineral. It is also a common accessory mineral in the Earth's crust and builds vast chalk and limestone sediments. These carbonate rocks are porous and prone to chemical compaction because of the relatively high reactivity of calcite in contact with percolating fluids[10]. The reactivity of calcite in the confined interfacial regions may significantly contribute to either rock consolidation by cementation[11,12] or weakening by brittle and plastic deformation[13,14]. Although recent studies related to carbonate-fluid interactions have suggested that surface forces may influence the mechanical strength of carbonate rocks [12,15-18], direct measurements of the forces between calcite surfaces in aqueous solutions varying in ionic strength and composition are limited[16,19-22].

Salinity has a pronounced effect on nm-range forces between two calcite surfaces. Strong repulsive hydration forces due to hydration of the highly hydrophilic calcite surface have been recently measured both in water and in electrolyte solutions[16,20] and found to significantly exceed the electrical double layer repulsion. The onset and magnitude of the hydration forces have largely depended on the electrolyte concentration, with smaller onsets at higher concentrations[20]. The collapse of surface hydration layers at high ionic strengths (>0.1 M NaCl) and electrostatic attraction due to ion correlation may be the two dominant mechanisms that facilitate adhesion between calcite surfaces, as suggested by Javadi and Røyne [22]. Adhesive forces between two calcite surfaces have been also measured at strongly alkaline conditions (pH =12, 0.12 M), pointing to weaker repulsion at low calcite zeta potentials[19].

Salinity also influences calcite reactivity. The salinity of pore waters that saturate sedimentary rocks can vary within 5 orders of magnitude, reaching as high as 0.3 kg/L (~5 M NaCl) of dissolved solids[23]. Mixing, migration of these waters, and anthropogenic injection of fluids into carbonate rocks can lead to temporary disequilibrium conditions and activation of growth and dissolution processes. Calcite solubility and growth kinetics in salt solutions are mainly affected by changes in ionic strength, ion hydration, ion pairing, and the common ion effect[24,25]. As the solution ionic strength increases, the activity of species that build the solid phase decreases in the solution, causing a higher solubility of calcite[26]. The dissolution rate of calcite has been found to increase at higher ionic strengths (>1 mM), owing to the ion-specific changes in $Ca^{2+}$ solvation and the resulting disruption of calcite surface hydration layers[27,28]. Background ions also have a profound impact on $CaCO_3$ nucleation, since they affect the dehydration energy of emerging nucleation clusters and therefore lead to significant differences in the critical supersaturation required for nucleation[25]. Certain ions that can be incorporated into calcite lattice (e.g. $Mg^{2+}$), will additionally modify the calcite solubility due to the impurity effect[29].

Spatial confinement can have a manifold effect on calcite reactivity. Ion depletion and reduced ion mobility in pores make nucleation events less probable[30,31], which increases induction times for crystallization. Single, μm-sized crystals grown in confinement display diffusion-limited rim topographies[7,32]. At the nanoscale, confinement effects may be even more pronounced: If the pore dimensions are smaller than the critical nuclei size, the surface free energy barrier may prevent nucleation altogether[33]. Nanoporous materials may selectively control the growth of different $CaCO_3$ polymorphs[34], and pore size-related changes in ion distribution near charged surfaces may promote growth of otherwise unstable phases[35]. Interestingly, Stephens, et al. [36] have recently observed that even μm-range confinement can promote nucleation of amorphous calcium carbonate (ACC) over the more stable $CaCO_3$ polymorphs. The authors suggested that despite the lower surface free energy of ACC with respect to calcite, its stabilization could not have been of thermodynamic origin since the bulk free energy gain on recrystallization into calcite dominated for surface separations larger than a few nm. They therefore attributed the stabilization of ACC to kinetic effects related to restricted ion transport in the confined solution[36].

It is not clear how changes in salinity affect interactions between confined calcite surfaces. On the one hand, attractive, short range forces between calcite surfaces should dominate in concentrated electrolyte solutions, leading to strengthening of interfaces at grain contacts[22]. On the other hand, calcite surfaces become more soluble and reactive in high salinity solutions, which could make the grain contacts weaker. The surface reactivity of confined calcite interfaces can lead to transport-dependent recrystallization processes and major increases in surface roughness[21]. Roughness and crystal growth may in turn give rise to very strong repulsive forces linked with the force of crystallization[3] and nanoscale asperity deformation[37,38]. Even in bulk solution conditions, $CaCO_3$ has been demonstrated to follow a non-classical crystallization pathway, with dense hydrated prenucleation clusters identified at initial crystallization stages[39,40]. These clusters may be present also at undersaturated solution conditions[41]. If such a dense phase persists between confined surfaces, it may significantly affect the nm-range forces between calcite surfaces.

Measurements of surface forces and reactivity of confined mineral interfaces remains a challenge since few methods are able to follow both the forces and topographical evolution *in situ* with sufficient resolution. In this work, we used the surface forces apparatus (SFA)[42,43] coupled with multiple beam interferometry (MBI)[44-47] to measure both the nm-range forces between two rough, polycrystalline calcite surfaces and their surface reactivity in confinement, the latter with μm-scale resolution. We performed the measurements in NaCl, $CaCl_2$ and $MgCl_2$ electrolyte solutions with ionic strength ranging from 10 mM to 1 M. The geometry of two contacting surfaces in our SFA experiment resembles an open slit pore with nm to sub-μm distance between the two opposing walls in the contact area with a typical radius of 50-100 μm. In contrast to a standard Atomic Force Microscopy (AFM) experiment with nm-sized contact areas, such a large contact area significantly affects the transport of ionic species and thus surface reactivity[21]. As such, our sample setup is relevant for confined interfaces not only in geological environments but also in granular, mineral-based materials.

# Results and Discussion

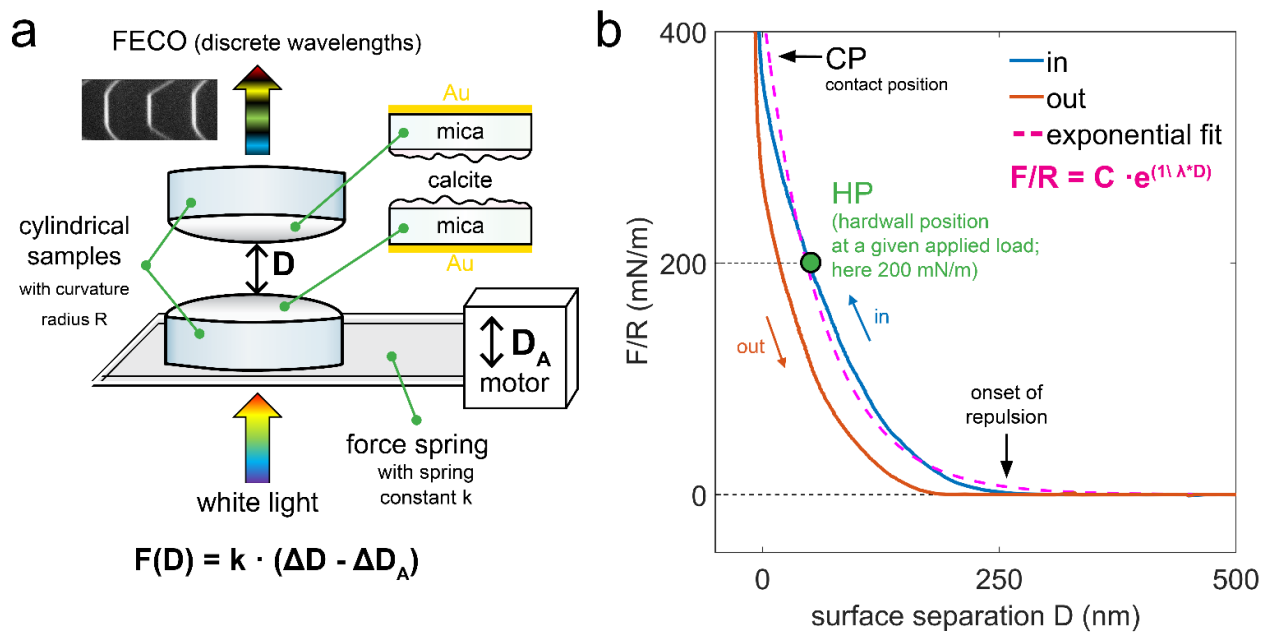

**Figure 1.** *a) Schematic representation of the SFA setup with two calcite surfaces glued to two crossed cylindrical disks with radius of curvature R. Surfaces are mounted on a force measuring spring, which is driven up and down at a constant velocity by a distance $D_A$. The actual distance between the surfaces $D_M$ is measured by optical MBI technique: fringes of equal chromatic order (FECO) form after passing white light through semi-reflective samples with nm-thick Au mirrors. Mica is used as a support to deposit calcite and Au films; b) Representative SFA force (F) measurement showing F normalized with R as a function of surface separation D on approach (in) and on retraction (out). Important parameters were marked on the plot. CP is defined as the distance at which the separation between the surfaces does not decrease significantly despite the continued loading. Exponential fit to the force-distance curve on approach is used to determine magnitude and range of repulsion by using exponential decay length λ; C is a fitting coefficient.*

We used the SFA to measure forces ($F$) as a function of surface separation ($D$) between rough and polycrystalline calcite surfaces in three calcite-saturated electrolyte solutions: NaCl, $CaCl_2$ and $MgCl_2$ with ionic strengths ($IS$) of 0.01, 0.1 and 1 M. During the SFA experiments, we performed repeated loading-unloading cycles, in which a bottom calcite surface (mounted on a force measuring spring) was approached towards and retracted from a top calcite surface at a constant velocity (ranging from 1 to 10 nm/s). In the SFA, the distance between surfaces is measured with an optical multiple beam interferometry (MBI) technique: the wavelength positions of a set of fringes of equal chromatic order (FECO; that result from light transmission through two semi-transparent samples) correspond to a given separation between surfaces. Positions of the FECO fringes are also sensitive to refractive indices of layered samples. The SFA setup and the most important parameters of measured force curves are shown in **Figure 1**. The details of SFA and MBI techniques and preparation of calcite samples for the SFA have been previously described[21,43-45,48].

We observed a clear and reproducible pattern of events during our SFA experiments: (1) forces between two calcite surfaces were initially monotonically repulsive, with no resolved attraction or adhesion in any of the solutions, even for the smoothest calcite surfaces; (2) calcite surfaces initially dissolved in contact with all solutions, and in most of the experiments, these dissolution periods were followed by major precipitation events. During these events, distinct *precipitation fronts* (PF) passed through the observed contact areas; and (3) after the passage of the *precipitation fronts*, the magnitude and onset (taken as the distance at which the force becomes of measurable magnitude; **Figure 1B**) of the repulsive forces substantially increased, to the extent that it could not be explained by roughening or damage of the calcite samples. In the following sections we first discuss the origin of the moderately repulsive forces before PFs, then we characterize PFs, and last, we discuss the long-range repulsive forces measured after PFs.

## Calcite surfaces

In line with previous findings[21,49], X-ray Diffraction (XRD) indicated that all the $CaCO_3$ films prepared by Atomic Layer Deposition (ALD) were composed of calcite (**Figure S1**). Two sets of ALD calcite surfaces were used for the SFA measurements. Although the ALD deposition parameters were kept constant **(Table S1)**, these films differed in morphology and initial roughness (**Figures S2-S5**) due to high sensitivity of the ~8 h deposition process to the deposition parameters and substrate characteristics[49]. Set 1 surfaces were composed of small crystals (50 – 100 nm), with a relatively high amount of much larger (~ 1 μm), polycrystalline aggregates particles on the surfaces (**Figures S2, S4**). The root-mean-square (rms) roughness of the set 1 films varied by almost 2 orders of magnitude due to the random distribution of the large aggregates (**Figure S5, Figure S2C**). Set 2 surfaces were more homogenous with larger, platy crystals (>200 nm) and continuous coverage of smaller crystals (50 – 100 nm; **Figures 2B-C, S3, S4**), and an initial average rms value of 4.3 ± 0.8 nm (as measured in 3 positions on a sample, scan size 15x15 μm$^2$; **Figure S5**). Despite using calcite-saturated electrolyte solutions in our experiments, we observed a minor initial dissolution of all the calcite films. This was mainly related to: a) disequilibrium morphology of calcite crystals grown by ALD from vapor phase, with the possible presence of high-energy crystal faces, as reported previously[21,49]; b) large roughness of the substrate composed of nm-sized crystals with large surface to volume ratio[50]; and c) uncontrolled pCO$_2$ conditions during saturating the salt solutions with calcite, and during the experiments. Nevertheless, our calcite films remained intact and continuous throughout the SFA experiments, even when 1 M *IS* solutions were used. We additionally measured the evolution of surface roughness for unconfined ALD calcite surfaces (set 3) using the AFM in 0.01 and 1M *IS* NaCl, CaCl$_2$ and MgCl$_2$ solutions, presaturated with calcite **(Figures S6, S7)**. We observed minor changes in surface roughness within several hours (Δ $rms$ < 3 nm). Only the most concentrated (1 M) NaCl solutions induced substantial dissolution of the calcite films, with μm-sized dissolution pits developing on the surfaces within the first 4 h (**Figure S7**).

## Origin of the repulsive forces before the precipitation fronts

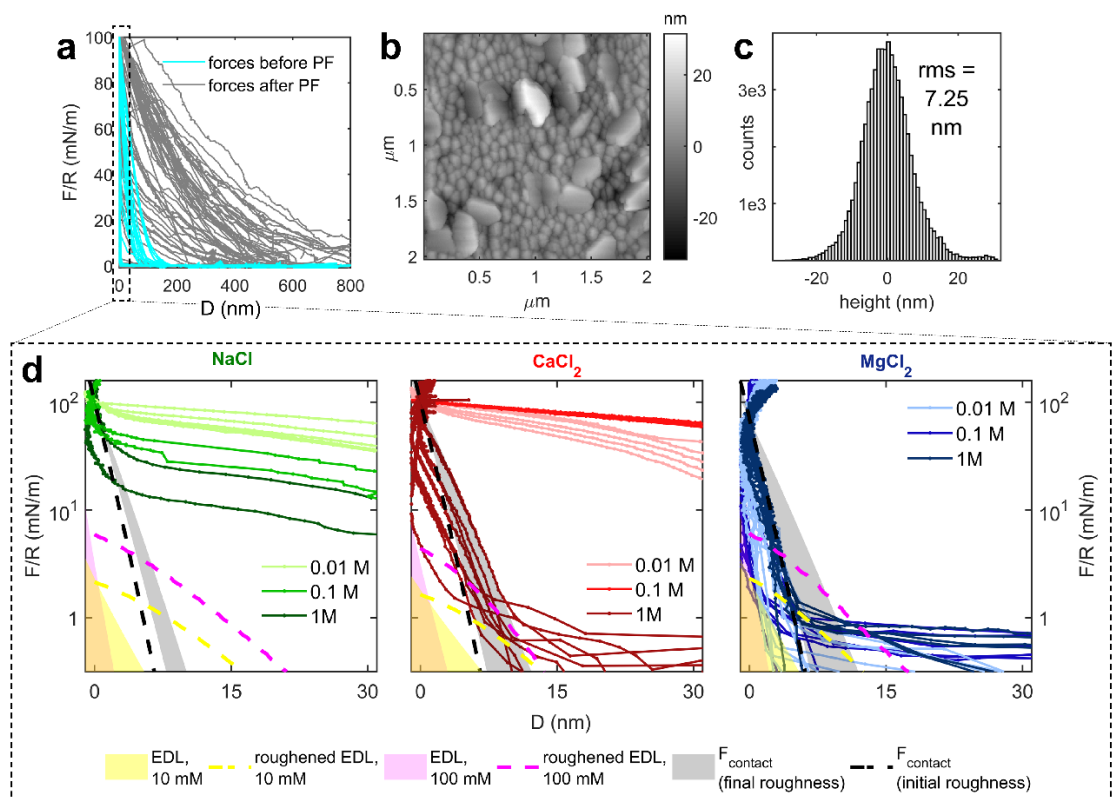

**Figure 2.** *SFA force measurements between two rough polycrystalline set 2 calcite surfaces. **a)** Summary of forces (F/R; normalized with radius of curvature R) measured as a function of surface separation (D) before and after passage of precipitation fronts (PF). Note much smaller range and magnitude of the repulsion measured before PFs; **b)** Representative AFM height map for the used calcite surfaces measured before the SFA experiments; **c)** Histogram of surface heights corresponding to B; **d)** Measured force curves (whole drawn lines) between two calcite surfaces (set 2) before PF events (cyan force curves in subplot A) in NaCl, CaCl$_2$ and MgCl$_2$ electrolyte solutions with IS = 0.01 to 1M, along with the modelled electrical double layer (EDL) repulsion and two roughness contributions (F$_{contact}$ and roughened EDL).*

In each SFA experiment, we measured forces as a function of surface separation for the same µm-sized contact for 2 days (set 1) or 1 day (set 2). In this section, we only discuss the moderately repulsive forces measured in the initial stages of the experiments, before PFs **(Figures 2A, D)**. The forces measured before PFs were monotonically repulsive and we could not resolve any attractive or adhesive forces, even for the solutions with high ionic strength. We semi-quantified the magnitude and onset of the repulsion using the decay length $\lambda$ of the exponential fit to the force-distance curves measured on approach[21,38,51] (**Figure 1B**).

The decay lengths of the force curves measured on approach before PFs were of comparable order for the two sets of surfaces (set 1: 6 nm < $\lambda$ < 35 nm; set 2: 1 nm < $\lambda$ < 65 nm). **Figure 2D** shows all force curves obtained on approach for set 2 surfaces, measured before PFs in NaCl, $CaCl_2$ and $MgCl_2$ solutions. Set 2 surfaces were smoother than set 1 surfaces and small loads were sufficient to reach the contact position (CP; defined as the distance at which the separation between the surfaces does not decrease significantly despite continued loading; **Figure 1B**). Flattening of the FECO fringes[48] observed at CP additionally indicated that surfaces were in a very close proximity: the minimum separations between surfaces were initially <10 nm over the whole nominal contact areas (~ 100 µm) for most of the set 2 experiments (**Figures 2D; S10**). For these experiments, we observed major differences in the range of repulsion in different solutions. We measured relatively long-range repulsion (with onsets at > 100 nm and 16 nm < $\lambda$ < 65 nm) in NaCl and in 0.01 and 0.1 M $CaCl_2$ solutions **(Figure 2D)**. Shorter-range repulsion (with onsets < 15 nm and 1 nm < $\lambda$ < 6 nm) was measured in $MgCl_2$ and 1M $CaCl_2$ **(Figure 2D)**.

Set 1 surfaces were much rougher than set 2 surfaces and large, µm-sized asperities (**Figure S2C**) prevented the surfaces from reaching CP. These asperities gave rise to comparable decay lengths for the set 1 and set 2 surfaces, because they acted as discrete hard walls at large separations (< 1 µm; **Figure S10**), and very high loads had to be applied to move the surfaces further in (as these large asperities plastically deformed); nm-range separations, at which surface forces operate, were thus only accessible for the highest asperities in the contact region. This explains why we did not resolve any major differences related to ionic strength or solution composition for the rougher set 1 surfaces.

As we observed major differences between decay lengths of repulsive force curves measured for the smoother set 2 surfaces in different electrolyte solutions (long-range repulsion in in NaCl and in 0.01 and 0.1 M $CaCl_2$ and short-range repulsion in $MgCl_2$ and 1M $CaCl_2$), we modelled which type of forces could explain the variation in the range of repulsion (**Figure 2D**). We considered possible contributions of: Van der Waals (VdW) forces, hydration forces, roughness and electric double layer (EDL) forces. As explained below, we neglected the attractive VdW forces and repulsive hydration forces. For the remaining two repulsive terms, we treat the effect of roughness explicitly and show that neither roughness nor EDL forces were sufficient to explain the measured variation in the range and magnitude of the repulsion. We suggest that the long-range repulsion was related to nucleation in the solution confined between the surfaces, even before the observable PFs.

In general, DLVO theory predicts relatively strong Van der Waals (VdW) attractive forces (Hamaker constant of $1.44 \cdot 10^{-20}$ J for two calcite surfaces across water[52]) to act between calcite surfaces. This attraction is however weakened by hydration of hydrophilic calcite surface, which gives rise to structural repulsive forces[16,20,53]. Therefore, in previous experiments, adhesive forces between two calcite surfaces have been measured only in high electrolyte concentrations (>0.1 M)[22] or at high pH (12)[19], and attributed to the collapse of the surface hydration layer and strengthening of ion-ion correlation forces[22], or EDL screening at low calcite zeta potentials[19]. Since we did not resolve any attraction nor adhesion, we did not include the VdW contribution.

It is likely that hydration contributes to the repulsive forces that we measured. Hydration repulsion is typically a nm-range, monotonically decaying force, generally described with decay lengths <2 nm for smooth surfaces[53]. Due to the unconventional Stern layer of calcite, where water molecules are directly adsorbed to the surface[54,55], both strong primary hydration (due to these adsorbed water layers) and secondary hydration related to surface-adsorbed cations are expected to act between calcite surfaces[53]. Since there are only semi-empirical expressions accessible to account for hydration repulsion, which do not include ion concentrations directly[53], and since hydration will be greatly affected by surface-specific ion binding, we did not include this interaction in our modelling. However, we may expect hydration repulsion of a weaker magnitude in high electrolyte concentrations[20], a trend that roughly follows EDL interactions for calcite surfaces.

The magnitude and range of EDL repulsive forces between similar surfaces are related to the surface charge and ionic strength of the solution. A precise determination of the EDL contribution for two calcite surfaces is challenging because of the large variation in reported calcite zeta potentials and their sensitivity to pCO$_2$ and solution composition[56-60], as well as few reported values for the calcite surface charge regulation parameters[20], which cannot be measured using rough and reactive calcite surfaces. Therefore, we chose to consider the possible range of EDL forces corresponding to the absolute zeta potential values of 5 to 30 mV, typically reported for calcite at pH ~8-9[60]. The EDL force contribution was calculated using a linearized Poisson-Boltzmann equation and calcite charge regulation parameter estimated by Diao and Espinosa-Marzal [20]. Details of the calculations are outlined in the Supplementary Information. The calculated Debye length of our electrolyte solutions varied between 3.0 - 4.3 nm for the 0.01 M electrolytes and 1.0 - 1.4 nm for the 0.1M electrolytes (as calculated including $Ca^{2+}$, $CO_3^{2-}$, $HCO_3^-$ species due to calcite dissolution upon pre-saturation, using PhreeqC). At 1 M, DLVO theory breaks down and EDL forces should be negligible due to strong ionic screening[61]. As such, the maximum range of the theoretically calculated EDL is ~15-30 nm for smooth calcite surfaces in our most dilute 2:1 electrolytes and at the highest surface charge (30 mV; **Figure S8**).

Surface roughness contributes to the measured repulsion in two ways. First, roughness produces repulsive mechanical effects due to plastic and elastic deformation of surface asperities on loading, the magnitude of which generally increases exponentially (for surfaces with random distribution of surface heights such as our set 2 calcite surfaces; **Figure 2C**) with decreasing surface separation[37,38]. The onset of this repulsion is related to the distance at which the first large asperities come into contact, roughly at distances smaller than 3 times the rms roughness of the surfaces[38]. Second, roughness smears out any distance-dependent interaction potential due to variation of surface heights across the nominal contact area[38,62]. Due to disruption of ion layering near the surface and possible roughness-related variations of surface charge, these roughness effects may extend over the full width of the EDL[63].

To account for these two roughness contributions, we used the model proposed by Parsons, et al. [38]. We estimated the first roughness contribution due to elastic or plastic asperity deformation from $Eq$. 16 in Parsons, et al. [38] ($F_{contact}$; **Figure 2D**) This contribution is based on rms roughness (measured with AFM for the ALD surfaces before ($F_{contact}$ initial) and after ($F_{contact}$ final) the SFA experiment, at three random positions for each surface; scan size 15x15 µm$^2$), average asperity radius (approximated from the AFM maps by measuring radii of areas above a height threshold of 70 %), and the Young's modulus and Poisson ratio of calcite. The second roughness contribution due to variation of surface heights across the nominal contact area was calculated using $Eq$. 7 in Parsons, et al. [38] (roughened EDL; **Figure 2D**) This contribution was modelled by averaging the theoretical EDL force (calculated with $Eq$. S2 for $\psi_0$ = -30 mV) for smooth calcite surfaces over the distribution of surface heights measured with the AFM for each surface (scan size 15x15 µm$^2$). The Derjaguin approximation was used to relate the calculated roughness-related interaction energy to the force acting between two cylindrical SFA samples (see $Eq$. 1 in Parsons, et al. [38]).

The results of force modelling **(Figure 2D)** indicated that the EDL and roughness force contributions are not sufficient to explain the long-range repulsion measured in NaCl and in 0.01 and 0.1 M CaCl$_2$ solutions: (1) EDL repulsive forces calculated for smooth calcite surfaces can be of measurable magnitude at separations < 10 nm. The onset of EDL forces may be larger (separations > 15 nm) if we consider roughness-averaged EDL forces (roughened EDL). However, even the roughened EDL cannot explain the measured long-range repulsion with onsets > 100 nm (and 16 nm < λ < 65 nm). EDL forces may significantly contribute to the short-range repulsion measured in MgCl$_2$ and 1 M CaCl$_2$, however it is not possible to precisely distinguish it from the roughness $F_{contact}$ contribution, which becomes of significant magnitude at comparable separations; (2) Roughness contribution due to asperity deformation ($F_{contact}$) can explain the high-magnitude, short-range repulsion with onsets below 15 nm (and 1 nm < λ < 6 nm) measured for the experiments in MgCl$_2$ and 1M CaCl$_2$ solutions. The magnitude and range of the experimentally measured repulsion in these experiments corresponds very well to the $F_{contact}$ force that was calculated using the roughness parameters measured for the probed calcite surfaces with the AFM. Since the roughness of set 2 surfaces was homogeneous and comparable for all samples (**Figure S5C, D**), repulsive forces due to surface roughness cannot explain the long-range repulsive forces measured in NaCl and 0.01 and 0.1 M $CaCl_2$ solutions.

What is the potential origin of the long-range repulsion that we measured before the observable PFs? If roughness was to explain the long-range repulsion in NaCl and 0.01 and 0.1 M $CaCl_2$, then the rms roughness of these surfaces (according to *Eq. 16* in Parsons, et al. [38]), would need to be one order of magnitude higher than measured with the AFM (up to rms ~ 100 nm for 0.01 M NaCl). We did not measure such a large roughness for any of the calcite surfaces used in the experiments in which the long-range repulsive forces were measured. Since it is not possible to precisely locate the contact region on the samples after the SFA experiments, we measured the surface roughness in 3 random locations on each sample with the AFM. Interestingly, we did not find any major increase in the sample roughness even after the major precipitation fronts (PFs; **Figure S5C, D**). Since we also investigated each sample with the Scanning Electron Microscopy (SEM; **Figures S2, S3**), it is unlikely that we overlooked features on the surface that could give rise to such large roughness.

Possible sample damage during the experiment, such as large calcite particles (~0.1 μm) breaking off and becoming trapped between the surfaces, could potentially explain the long-range repulsion with onsets >100 nm. Although we do not observe any loose particles in the camera or any major irregularities in FECO fringes, the size of such particles could have been below the μm-range resolution of the FECO and our camera. However, with large particles trapped between the surfaces we would not observe a pronounced flattening of the surfaces (due to elastic deformation of the glue) at the contact position, as the pressure would be concentrated on the discrete asperity contacts that are much smaller than the nominal contact area in the SFA. Flattening was observed for all set 2 experiments before the PFs.

No long-range repulsion was measured for any experiments in $MgCl_2$ and 1 M $CaCl_2$ solutions. We observed little reactivity of calcite films in $MgCl_2$ solutions, for which distinct PFs appeared only in two experiments (0.1 M and 1 M $MgCl_2$, set 1) after 12 and 15 h, respectively. Calcite surfaces in 1 M $CaCl_2$ were also the least reactive in comparison with 0.01 and 0.1 M $CaCl_2$ experiments. This suggests that the initial long-range repulsion measured in NaCl and $CaCl_2$ (0.01 and 0.1 M) was in some way related to the reactivity of the calcite surfaces.

Ruling out changes in calcite surface roughness, EDL forces and surface damage as explanations for the initial presence of long-range and high-magnitude repulsive forces leaves us to consider the properties of the solution confined between two calcite surfaces. In the following, we will show that the fluid compositions with presence of long-range and high-magnitude repulsive forces are the compositions where we later observed distinct precipitation fronts. As such, the long-range repulsion measured before the PFs was likely related to nucleation of an amorphous $CaCO_3$ phase between the calcite surfaces.

Precipitation fronts

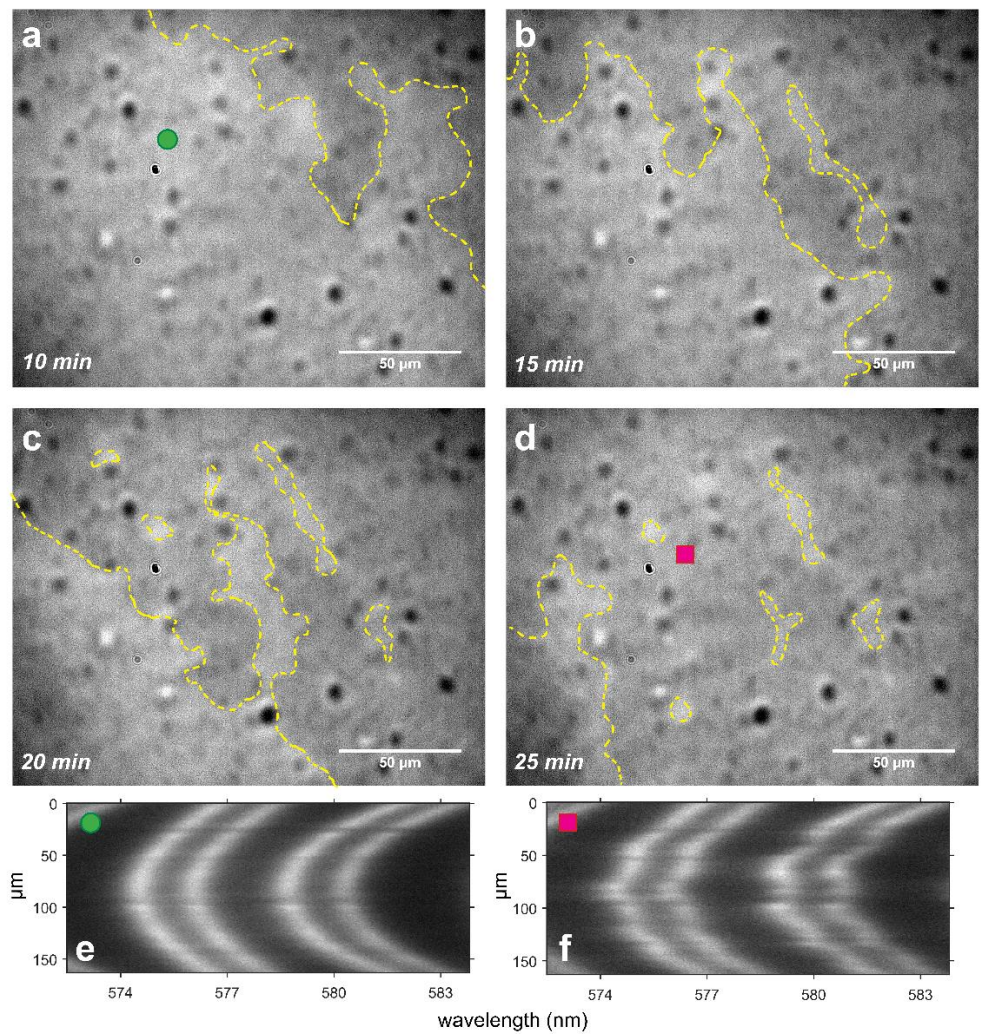

**Figure 3.** *a-d) Spreading of precipitation front (PF) between two calcite surfaces in the SFA, indicated by a darker color of the precipitate (set 1, 0.1 M NaCl experiment; supplementary movie M2); e) FECO fringes before PF (surfaces out of contact); f) FECO fringes after PF (surfaces out of contact). The center of the contact area established in the SFA (corresponding to the shown FECO fringes) is approximately indicated with a green or a pink symbol.*

In almost all SFA experiments, we observed distinct *precipitation fronts* (PFs) passing through the imaged contact regions (**Figure 3**). PFs were manifested by fingering growth of a darker region between the two calcite surfaces, spreading into the spherical contact area from outside of the contact with velocities ranging from ~10 to ~500 nm/s **(Figure 3A-D, supplementary movies M1-M14).** The PFs could be identified in the camera by a change in intensity of the light (transmitted through two semi-transparent calcite surfaces in the SFA) from brighter to darker, which could be caused both by a change in separation between surfaces and a change of the refractive index of the solution trapped between the surfaces. The PFs could be also identified from the changing position and shape of the FECO fringes[45,48], which were gradually losing their resolution (became wider) and became very irregular **(Figure 3E-F)**. Whenever PFs reached the contact region, positions of FECO fringes shifted to wavelengths corresponding to larger separations (as determined from experiments in which surfaces were kept at a fixed separation under constant load during PF). The irregular shape of the FECO fringes could indicate changes in surface topography or uneven refractive index (and thus uneven density) of the solution confined between the surfaces. Since we did not observe any major changes in calcite surfaces topography after the SFA experiments (**Figure S5**), the irregularity of the FECO fringes likely indicated variations in the density of the solution confined in the contact region. Based on these observations, we interpret these fronts to represent precipitation events.

We had no possibility to directly identify the material precipitating between the calcite surfaces, but since we could not resolve any distinct particles or crystals, the precipitate was most likely of poorly crystalline (<<µm) or amorphous nature. Given the chemical composition of the surfaces and solutions, it is unlikely that any mineral phase other than $CaCO_3$ would precipitate. We did not observe any distinct precipitate after the SFA experiments with AFM or SEM (**Figures S2, S3**), suggesting that the precipitate remained in the solution and was lost on disassembling the SFA surfaces. Observations in the camera, when repeatedly approaching and separating the surfaces, suggested that the new phase was a dense liquid-like suspension of the amorphous or nanocrystalline precipitate that could flow into and out of the contact region (**supplementary movie M15, Figure 4A-D).** Since $CaCO_3$ has been shown to follow a non-classical crystallization pathway, with a dense liquid-like phase separating from the solution even in undersaturated conditions[25,39-41,64], we suggest that the observed precipitate was an amorphous, strongly hydrated phase of $CaCO_3$.

The precipitate formed in the solution trapped between the surfaces, and was not strongly attached to the calcite surfaces. This was manifested in several ways: (1) we could visibly displace most of the newly-formed precipitate from the contact area when we approached the surfaces manually at very high loads >> 1 MPa (using the manual SFA micrometer control[43]; **Figure 4A-D**); (2) by forcing the surfaces into contact manually at very high loads (>> 1 MPa), it was possible to reach the initial CP, which indicated no major change (< 10 nm) of the calcite layer thickness in the contact region (**Figure S9**); (3) we observed the changes in appearance of the FECO fringes: when the precipitate was present between the surfaces, the FECO fringes were very irregular; but when we squeezed the precipitate out the contact, the FECO fringes became very regular (**Figure 3E-F**); (4) there was almost no change in calcite roughness measured at the end of the experiments, especially for the more uniform set 2 surfaces (**Figure S5**). Unless we applied very high loads manually to squeeze the precipitate out of the contact, it remained between the surfaces until the end of the experiment (at applied loads < 1 N/m ~ 0.5 MPa; **Figure S10**).

As the position of the FECO fringes depends on the thickness and refractive index of each layer comprising the sample (in our case *Au-mica-calcite-solution-calcite-mica-Au*), it should be possible to estimate the thickness of the precipitate in the contact region. We used an exemplary force-distance measurement after PF, in which the HP position (defined as the separation at a given applied load; **Figure 1B**) was not changing significantly upon further increase in applied load. We assumed that this HP corresponded to the equilibrium thickness of the precipitate in the contact region at the given load (**Figure S12C-D**; set 2, 0.1 M $CaCl_2$). Assuming that the solution had the refractive index of water ($n_{H_2O}$), the minimum separation between the surfaces after the PF at applied load of ~200 mN/m was ~ 650 nm (**Figure S12B**). If there was a large difference between $n_{H_2O}$ and $n_{precipitate}$, then this minimum separation could be largely overestimated. The dense precipitate likely had a higher $n$ similar to a strongly hydrated ACC phase ($n_{ACC}$ ~1.5)[65]. Using $n_{ACC}$ ~1.5, the minimum separation between surfaces decreases to ~ 500 nm. Even if we used $n_{calcite}$ ~1.65, the separation is > 400 nm. This shows that the precipitate prevented the surfaces from coming into contact at moderate applied loads. A simple calculation, assuming a density of hydrated ACC ($\rho$~2.6 g/cm$^3$)[64], indicates that the precipitate could not have filled the entire volume between the surfaces (taking into account the maximum amount of Ca$^{2+}$ from dissolving ALD calcite films, and the Ca$^{2+}$ already present in the presaturated electrolyte solution), meaning that the precipitate must have been present as discontinuous domains or been of much lower density.

We never saw PFs initiating in the contact region established in the SFA, but rather propagating into the contact from outside the field of view. The region visible in our camera covers ~200 x 150 µm. This means that the observed PFs were initiated at distances >100 µm away from the location of the minimum surface separation. Due to the cylindrical geometry of our samples ($R = 0.02\ m$), the surface separation ($D$) varies as a function of distance from the contact position ($x$) and can be approximated as $D = R - \sqrt{R^2 - x^2}$ [36]. The separation between two surfaces 100 µm away from the contact position is <0.3 µm. Earlier reports suggest that the influence of confinement on calcite crystallization can be present for surface separations <10 µm[36]. Then the 'confined' area (with radius of ~600 µm) in our SFA setup is 40 times larger than the nominal contact area. The largest separation between the surfaces is ~0.7 mm at the edges of the samples.

It is puzzling that the precipitation occurred in the solution confined between two surfaces and not by heterogenous nucleation onto the rough calcite surfaces. The rough surfaces contain plenty of favorable nucleation sites where the contact between the precipitating phase and the substrate would be large. However, if the interaction between the nucleating particle and the surface is repulsive, precipitation is energetically favored in the bulk solution [33]. Moreover, the cryo-TEM observations of Pouget, et al. [39] showed that even in the presence of a favorable substrate for heterogenous nucleation, the first stable amorphous nanoparticles of CaCO$_3$ appeared in the solution before they aggregated on the substrate. If the precipitate that we observed was a highly hydrated amorphous phase of CaCO$_3$, and there was no driving force for its dehydration and homogenous nucleation in the confined solution, it is possible that there was also no driving force for its heterogenous nucleation onto the calcite surface (which was also hydrated[55]).

Interestingly, we observed nucleation of crystals in the contact region after the PF during repeated approaching and separation of the surfaces, when we moved the lower surface by means of the manual micrometer and repeatedly applied very high loads (> 1 MPa; **Figure 4E-I; supplementary movie M16**). Upon loading, most of the liquid-like precipitate was expelled from the contact region but it flowed back into the contact region on separation (**Figure 4A-D**). After several in-out runs, we observed ~5 µm particles appearing between surfaces. These particles were flat (separation between the surfaces was <1 µm when surfaces were in contact), loose (they were slightly changing position on the surface after each loading) and they first appeared when the surfaces were out of contact (**Figure 4E**). Although we did not directly identify these particles, they were very likely crystals of $CaCO_3$ (the particles scattered light, making the FECO fringes discontinuous). It is therefore possible that at the very small supersaturation of our solutions ($SI_{calcite}$ ~ 0), very high loads had to be applied to dehydrate the clustered ions[66] and trigger crystallization. Although the growth of the larger crystallites at the expense of the liquid-like precipitate suspension resembles an Ostwald ripening process[67], we never observed any spontaneous recrystallization of the liquid-like precipitate throughout the experiments (<25 h).

It is possible that the recrystallization of amorphous or nanocrystalline precipitate into larger crystals during the SFA experiments is hampered due to confinement. If we assume that at the low supersaturation of our solutions, the equilibrium size of $CaCO_3$ crystals growing between surfaces in the SFA is >1 µm, then stable crystals would not form unless they were able to displace the confining walls in order to reach that size. In the SFA setup, the top surface is fixed while the bottom surface is mounted on a force measuring spring with a spring constant $k$ = 2000 N/m. Even if we did not apply any load to the bottom surface, the growing crystal would have to overcome a confining pressure of the order of MPa to displace the bottom surface by a distance ($x$) of several nm (for a 1 µm² contact area), as estimated from $F = -kx$. At very low supersaturation, the crystallization pressure for calcite (calculated according to *Eq. 18* in Scherer [3] assuming equilibrium solubility of calcite 0.0130 g/L and solute concentration of 0.0131 g/L) should be of a similar MPa order (~1 MPa). Therefore, the presence of confining walls in our setup should not hamper the growth of µm-sized crystals (we have previously observed formation of µm-sized crystals in the SFA setup with much more soluble ALD calcite films grown at lower temperatures[21]) , making hindered dehydration is a more likely explanation for the lack of crystallization in confinement.

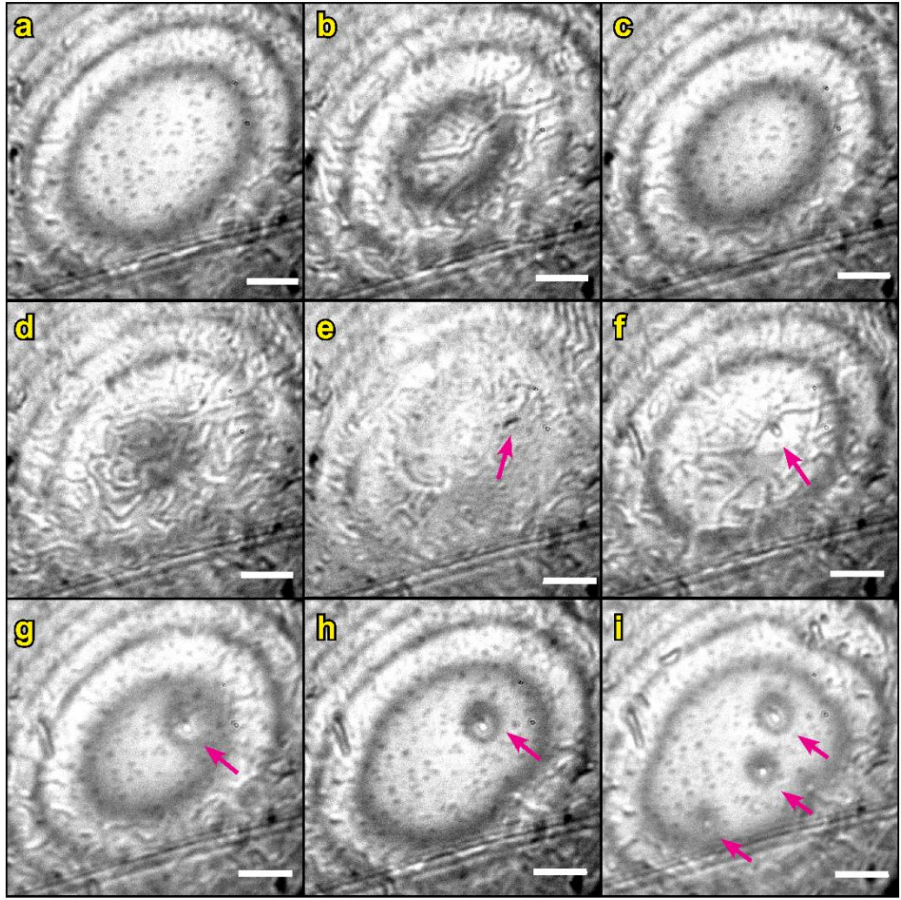

**Figure 4.** *Contact region between two calcite surfaces in the SFA after PF. Newton rings (interference fringes) connect regions of the same surface separation. The bright central Newton ring indicates a contact region of the smallest separation. The larger the diameter of the central Newton ring, the larger the nominal contact area; scale bar is 50 µm. Precipitate is identified as irregular, twisted features in the images. **a)** and **c)**: Precipitate is squeezed out of the contact region when the surfaces are approached manually at high loads. **b)** and **d)**: Precipitate flows back into the contact region upon surface separation. **e-i)**: After several loading-unloading cycles, µm-sized crystals grow between the surfaces (indicated with arrows), first when the surfaces are out of contact. Images **a** to **i** are a sequence in time (see also the supplementary movie M16).*

## Influence of electrolytes on reactivity

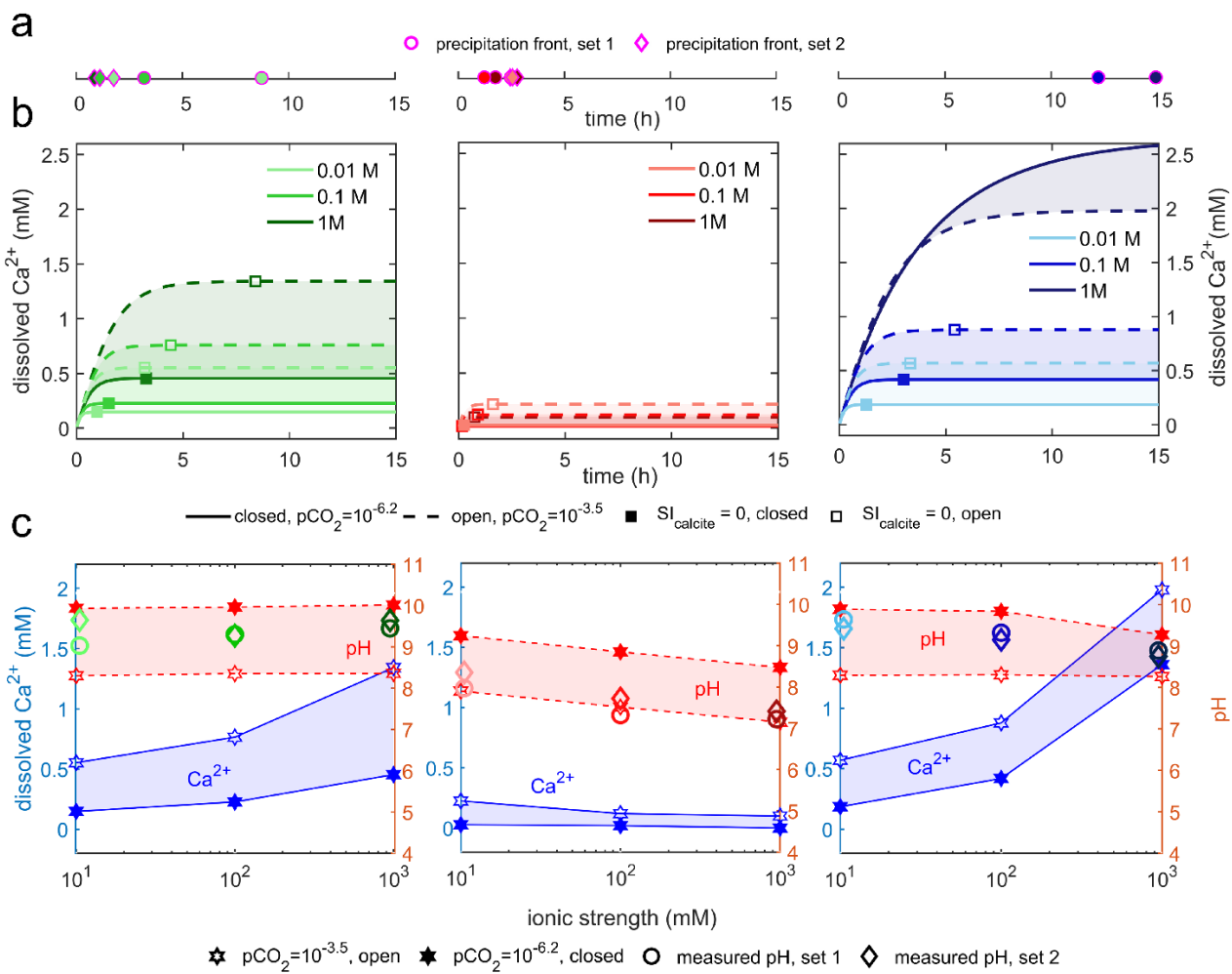

**Figure 5. *a*)** *Elapsed time before PFs measured for set 1 (○) and set 2 (◊) experiments;* ***b*)** *Dissolution kinetics of calcite modelled in PhreeqC in NaCl, CaCl$_2$ and MgCl$_2$ solutions with IS = 0.01, 0.1, or 1 M for closed (low pCO$_2$) and open (high pCO$_2$) systems. Squares (□) represent SI$_{calcite}$ = 0. Symbols on the top axis show.* ***c*)** *Parameters of solutions used in the SFA modelled in PhreeqC, showing a range of pH values and dissolved Ca$^{2+}$ for closed and open systems. ○ and ◊ symbols show the measured pH values for solutions used in set 1 and set 2 experiments.*

In **Figures 5A** and **6**, we show the timing of the PFs as a function of electrolyte composition. The time elapsed before the onset of the PFs was found to be largest for the MgCl$_2$ salts (PFs observed after 12 and 15 h only for two experiments in set 1: 0.1 and 1 M), and shortest for experiments in CaCl$_2$ (PFs were observed for all experiments within the initial 3h). PFs for experiments in NaCl occurred after 1 to 8 h. Calcite dissolution and precipitation kinetics are known to be affected by the presence of background ions due to: changes in ionic strength, ion-pair effects, ion solvation and a common-ion effect[24,25]. We used a simple model of calcite dissolution in PhreeqC to check if the differences in time elapsed before PFs could be related to ionic strength ($IS$) and composition of our solutions.

Using PhreeqC, we calculated the time and concentration of Ca$^{2+}$ required to reach supersaturation with respect to calcite (SI$_{calcite}$ > 0) in bulk electrolyte solutions that initially contained no dissolved CaCO$_3$. The calculations were performed using the rate for calcite dissolution defined in $'llnl.dat'$ database, assuming open (pCO$_2$ = 10$^{-3.5}$) or closed systems (pCO$_2$ = 10$^{-6.2}$). The rate equation is based on the model for calcite dissolution proposed by Plummer, et al. [68]. It has to be noted that the PhreeqC model is based on empirically determined surface to solution volume ratio and is only meant to show the relative effect of electrolyte composition on calcite precipitation. The modelling (**Figure 5B**) indicated: (1) low amounts of Ca$^{2+}$ (<0.5 mM) at saturation and fast equilibration (<2 h) for all CaCl$_2$ solutions due to the common-ion effect that decreases solubility of CaCO$_3$ in the presence of highly soluble CaCl$_2$; (2) larger dependence of $IS$ on the saturation level of NaCl and MgCl$_2$ electrolytes; and (3) higher solubility of calcite in MgCl$_2$ than in NaCl (especially at $IS$ = 1 M), due to the abundance of the MgHCO$_3^+$ ion pair that reduces the HCO$_3^-$ activity and shifts the calcite equilibrium. CP

Since the system used for PhreeqC modelling is different that the system that we have in our SFA experiments, we have to justify why we can make a comparison between these two. The solutions used in the SFA experiments were presaturated with calcite and had pH values characteristic for saturation under low pCO$_2$ conditions (**Figure 5C**; the concentration of Ca$^{2+}$, due to calcite dissolution, was modeled in PhreeqC). Since the solutions were saturated with respect to calcite, we should observe no dissolution of the calcite surfaces and thus no precipitation. However, our ALD surfaces were rough and due to high radius curvature of the nm-sized crystals, their solubility was likely higher than the solubility of the calcite powder used for saturating the electrolyte solutions[50]. Additionally, the ALD calcite films grown from the vapor-phase may be partially composed of crystals with high-energy faces that are much more soluble than the most stable {104} calcite faces[49]. These are the two main reasons why we observed initial dissolution of calcite surfaces in the SFA experiments. We previously argued that during PFs precipitate was formed in the solution between the calcite surfaces and not by heterogenous nucleation on the calcite surfaces. Therefore, the properties of solutions such as their ionic composition and *IS* are expected to influence the elapsed time before PFs. This is because the concentration of ionic $Ca^{2+}$, $CO_3^{2-}$ and $HCO_3^-$ that is necessary to reach supersaturation with respect to calcite depends on the electrolyte IS and composition (**Figure 5B**). Even though these ions were initially present in the presaturated solutions that we used in the SFA, we expect that the additional concentration of these ions needed to reach supersaturation depends on *IS* in a proportional manner. As such, the initial dissolution stage should be the longest and the onset of PFs should be the most delayed for the highest *IS* solutions for NaCl and $MgCl_2$ electrolytes. The opposite trend is expected for $CaCl_2$.

Comparison of the PFs onsets with the calculated theoretical equilibration times of the electrolyte solutions with calcite indicates that: (1) there was no clear correlation between the time onset of PFs and *IS* for any of the solutions, contrary to what was expected from calcite dissolution kinetics (**Figure 5A, B**); (2) the PFs for $MgCl_2$ occurred only after 12 and 15 h, and only in two experiments (set 1, 0.1 and 1M), although the modelled calcite dissolution rate was comparable for NaCl and $MgCl_2$ solutions at lower *IS* (0.01 and 0.1 M; **Figure 5A, B**). This suggests the importance of specific ion effects (e.g. ion solvation[25]), which are not considered in the PhreeqC modelling and are described in the following; (3) fast equilibration of calcite films and comparable time onsets of PFs for all experiments in $CaCl_2$ solutions (all within the initial 3h) agrees with the reduced calcite solubility in presence of $CaCl_2$.

We have previously shown that dissolution of calcite in the SFA contact region is affected by the contact roughness[21]. In the current study, the initial dissolution of calcite before the PFs was also correlated with the surface roughness (as estimated from the initial CP at the beginning of the experiments[21]; **Figure S11C**). However, we did not find any correlation between the onset of PFs and neither the amount of dissolved calcite before PFs, nor the initial contact roughness (**Figure S11**). This may be related to the fact that the surface separation where the PFs were initiated was of the order of several μm. There, the nm-scale surface roughness of the calcite films should not additionally influence the transport of solutes along the gap.

The lack of correlation between PF onset and ionic strength was then likely related to the location on the sample where the PFs were initiated and the distance they propagated before we identified them in the camera. The fact that we observed PFs at various stages after they initiated, is supported by the differences in front velocities (~10 to ~500 nm/s) and the spreading manner (with full or partial coverage within the observed area), both likely related to concentration gradients along the gap. As the separation between the surfaces continuously increases from the contact region towards the bulk solution, we expect both the differences in local dissolution rates of the calcite films and solute diffusion rate out of the gap to affect the time onset of PFs.

Although the effect of the *IS* on the onset of PFs may have been obscured in the SFA setup for the reasons outlined above, we still observed a significant effect of the cation type present in the solution. Since we did not identify any crystals and we suggest an amorphous nature of the observed precipitate, the effect of the cation may be related to the dehydration of $CaCO_3$ pre-nucleation clusters. It has been previously observed that there is a link between both the induction time and effective supersaturation for $CaCO_3$ nucleation and the way in which different background ions stabilize water in $CaCO_3$ pre-nucleation clusters. In general, it takes more energy to dehydrate smaller, multivalent cations. These cations structure water in their solvation shells to a higher extent, decrease water mobility, impede ion dehydration and hinder precipitation[25]. This is in line with our findings as the induction times for PFs are greatly delayed in presence of $Mg^{2+}$, which is the smallest and most hydrated cation in comparison with $Ca^{2+}$ and $Na^+$ (hydrated radius decreases in order $Mg^{2+}>Ca^{2+}>Na^+$). Whereas $Ca^{2+}$ is also relatively strongly hydrated, the effect of dehydration may be counteracted by the much lower solubility of calcite in $CaCl_2$ due to the common-ion effect. Interestingly, we have not observed any PFs in experiments performed in monoethylene glycol (MEG) over 3 days (Figure S13). MEG is known to delay the precipitation rate of $CaCO_3$, largely because of how the high viscosity of this solvent reduces the ion diffusivity[69,70].

## Long-range repulsive forces during and after precipitation

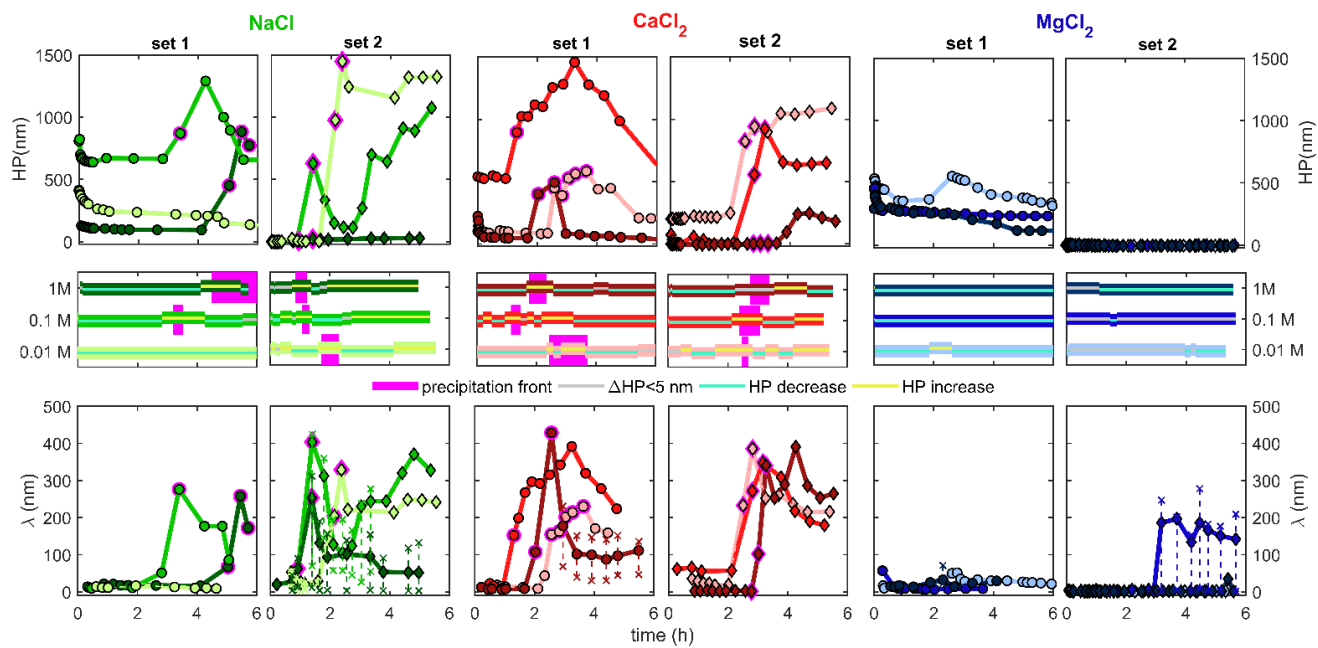

**Figure 6.** *Evolution of hardwall position (HP) and exponential decay length of repulsive forces (λ) with time for SFA force measurements between two calcite surfaces in NaCl, CaCl$_2$, and MgCl$_2$ solutions with IS = 0.01, 0.1 and 1 M for set 1 and set 2 experiments (colors correspond to ionic strength as shown in the middle panel). The middle panel shows periods of HP increase or decrease, and duration of the precipitation fronts (PF) in the observed region on calcite surfaces. Data points measured during PFs are outlined with magenta. Red x symbols show a range of possible decay lengths, whenever exponential fits were poorly fit to the force-distance curves.*

Both the hardwall positions (HP) and decay lengths (λ; see definitions in **Figure 1B**) increased significantly during and after the *precipitation front* (PF) events (**Figure 2A**). **Figure 6** shows the changes in hardwall position (HP) and the corresponding changes in decay length λ with time for all 18 experiments. The decay lengths measured after PFs (45 nm < λ < 400 nm) were many times larger than before PFs (1 nm < λ < 65 nm). Also shown is the duration of the PFs in the corresponding experiments. The location of HP for rough surfaces strongly depends on the applied load **(Figure S12C)**. We therefore determined the HP at the maximum load common to all measured force-distance curves in each experiment to clearly indicate major increases or decreases in HP as shown in **Figure 6**. Additional parameters of the measured force curves (such as the maximum applied load and minimum separations at the maximum applied load) are shown in **Figure S10**.

All experiments in which we observed PFs progressed in a similar manner (**Figure 6**): (1) the magnitude and range of repulsive forces were the smallest before PFs, and in all cases PFs were preceded by a period of calcite dissolution (indicated by decreases in HP << 500 nm). In $MgCl_2$ solutions, the calcite dissolution period was extended, with previously discussed late occurrences of PFs; (2) we subsequently observed rapid increases in both HP and decay length (λ); (3) these maxima correlated with the precipitation fronts (PFs) passing through the imaged contact areas; (4) after PFs, both HP and decay lengths (λ) decreased but were larger than at the beginning of experiments.

Whenever PFs reached the contact area, we measured a peak in repulsive forces (largest decay lengths). If PFs occurred when surfaces were kept in contact under constant applied load, we found surfaces to move out of contact by tens of nm (set 1: 0.1 and 1 M $MgCl_2$). This shows that the growing precipitate could act against loading and exert pressure on the confining walls. Although we could not determine whether the precipitate was of amorphous or poorly crystalline nature, this behavior shows that there was MPa-high 'crystallization' pressure[3] associated with the precipitation fronts.

After the precipitation had ceased in the contact region, both the magnitude and onset of repulsive forces and HP decreased again but they were always at least one order of magnitude larger than at the beginning of each experiment. **Figure 6** shows that the decay lengths of the repulsive force curves gradually decreased after PFs in most of the experiments. The decrease in decay length frequently corresponded to a decrease in HP. In some experiments, HP after the PFs gradually reached the initial HP position measured at the start of experiments (set 1: 0.1 M NaCl, 0.1 M and 1 M CaCl$_2$; set 2: 1 M NaCl; **Figures 6, S10**)[.] We interpret such evolution of the repulsive forces to be caused by progressive depletion of the precipitate from the contact region upon repeated loading-unloading cycles. Depletion of the precipitate in the contact area was observed in several cases in the camera (**supplementary movie M1**). We additionally measured large hystereses between the force-distance curves on approach and retraction that were not present or were very small before the PFs. Areas of these hystereses closely followed the trend shown for the decay length of the repulsive force curves in each experiment: they were the largest during PFs and decreased with time after PFs. The presence of these hystereses indicate that there was an energy cost related to the displacement of the precipitate from the contact region. As areas of the hystereses became smaller with time, this additionally shows that the precipitate was progressively squeezed out from the contact region. Based on the above observations we interpret that the long-range, monotonically decaying and hysteretic repulsion measured after PFs was related to the hydrodynamic drag caused by the high viscosity of the precipitate[71].

     The precipitate trapped between the calcite surfaces was likely denser and more viscous than the bulk solution. Although the exact viscosity of the precipitate was unknown (and the effective viscosity of the confined solution was influenced by the inhomogeneous distribution of the clustered precipitate in the gap), we expect a high viscosity suggested by the precipitate aggregation into discontinuous domains. High viscosity, increasing with the particle volume fraction, has been previously observed in colloidal suspensions of CaCO$_3$ nanoparticles[72]. Due to viscous forces, the precipitate would oppose the movement of the surfaces (similarly to what has been previously observed in SFA force measurements with non-adsorbing polymer melts[71]), giving rise to repulsive force on approach and hystereses between loading-unloading force curves. Assuming no-slip conditions, the hydrodynamic force $F_h$ is proportional to the movement velocity $v$ and fluid viscosity $\eta$, and can be estimated as $F_h = \frac{6\pi\eta R^2 v}{D}$ (for the crossed cylinder geometry of the SFA), where D is separation between the surfaces, R is

cylinder radius, and D << R[73]. We did not observe any correlation between the magnitude and onset of the measured repulsive force and the approach velocity in our experiments, something that should be present if hydrodynamic effects were at play. However, the range of the velocities that we used after PFs (~1 to 5 nm/s) could be insufficient to observe significant differences in the hydrodynamic contribution, especially with an inhomogeneous distribution of the viscous phase in the gap.

Long-range repulsion could additionally arise due to entropic, steric effects that are related to the confinement of the denser phase between the surfaces[61]. If the loading was too fast for the precipitate to be displaced from the gap, it could have become partially jammed between surfaces. Such trapped precipitate would oppose the surface movement either because energy was needed for its progressive dehydration or there was little available volume for its spatial rearrangement. We observed that after PFs and after several loading-unloading cycles some sort of CP was developing at large separations (hundreds of nm away from the initial CP, **Figure S12D**), where separation did not decrease despite further loading (e.g. set 2, 0.1 M $CaCl_2$ experiment, **Figure 6**). This reflected a high energy cost both to displace more precipitate from the gap and to further squeeze it in the gap (the range of applied loads that we used during the force measurements is plotted in **Figure S10**).

We have previously performed a series of similar SFA experiments, in which we used $CaCO_3$-presaturated solutions without added electrolytes[21]. We only observed major increases in the magnitude and onset of repulsive forces in a few experiments, in which the roughness of the surfaces was the smallest, and we have attributed this increased repulsion to the recrystallization of calcite surfaces. The findings of the current study, where we have performed a more thorough analysis of the roughness change after the experiments and we could observe reproducible PFs in almost all experiments, suggest that the increase in magnitude and range of the repulsive forces measured in the previous study was also likely related to $CaCO_3$ nucleation in the confined solution. The electrolyte solutions used in the current study speeded up the occurrence of PFs (either due faster dissolution of films in high-$IS$ NaCl solutions or decreased calcite solubility in $CaCl_2$ solutions). Thus, it was easier to trigger the PFs (even for the rougher contacts in the set 1 experiments), which required that a certain critical supersaturation with respect to calcite was attained in the gap between the surfaces.

# Conclusions

We showed that properties of the solution confined between two reactive calcite surfaces can affect interfacial forces even at µm-ranged surface separations. At low supersaturation with respect to calcite, we observed nucleation of an amorphous or poorly crystalline precipitate that formed in the confined solution. The precipitate, which was most likely a hydrated $CaCO_3$ phase, gave rise to long-range and high-magnitude repulsion acting between the calcite surfaces. These observations may have crucial consequences for the evolution of microstructure of both fluid-saturated rocks and mineral-based materials: (1) We measured the long-range repulsive forces at ionic strengths varying from 0.01 to 1 M. This shows that the strengthening of solid-solid contacts at high ionic strengths, as expected from the DLVO theory, can be counteracted by nucleation occurring in the solution confined between two solid surfaces; (2) The onsets of nucleation were influenced by ion specific effects to a higher extent than by the solution ionic strength, with $Mg^{2+}$ significantly delaying the nucleation; (3) The transport of reactants between mineral surfaces can be significantly slowed down in the presence of the dense precipitate that we observed, even at µm-range separations. Although it is generally expected that the transport of ionic species in confined solution should not be affected for separations larger than several nm, we showed a possible mechanism that can delay diffusion in relatively thick gaps. (4) Our measurements indicate that cementation of grain interfaces is not likely to proceed at low supersaturation conditions, as there exists an energy barrier for dehydration of the precipitate nucleating in confinement, even when the gaps between surfaces are µm-thick; (5) We showed that the occurrence of precipitation fronts in our system was correlated with the repulsive forces of the highest magnitudes. Therefore, the significant mechanical repulsion related to crystallization pressure can act on confining surfaces even when the nucleating phases are poorly crystalline or amorphous. Future work should involve more precise, *in situ* characterization of the nucleating phase.

# Methods

## Preparation and Characterization of Calcite Films

Thin (~200 nm), polycrystalline calcite films were grown at 300 °C by Atomic Layer Deposition (ALD) as described in Nilsen, et al. [49] using a F-120 Sat reactor from ASM Microchemistry. The detailed preparation of the calcite films deposited on mica substrates for the SFA has been explained in Dziadkowiec, et al. [21]. Because of substantial variation in roughness of ALD calcite films, we prepared 3 sets of surfaces, each in a separate ALD run. Sets 1 and 2 were used for the SFA measurements and set 3 was used for the AFM measurements in salt solutions. Detailed deposition and film parameters are provided in **Supplementary Information.** After the deposition, calcite films were kept in a vacuum-sealed desiccator.

X-ray diffraction (XRD) was used to identify the ALD-grown $CaCO_3$ phase on Au-coated glass slides (XRD peaks of mica substrate overlapped with the most intense calcite peak). We used Bruker AXS D8 Discover powder diffractometer in Bragg-Brentano configuration, equipped with a Lynxeye detector, using Cu K$\alpha$1 radiation and a Ge(111) monochromator.

Film morphology was observed with Scanning Electron Microscopy (SEM), using Hitachi SU5000 FE-SEM in secondary electrons (SE) mode (15 kV). The samples were coated with ~3 nm of Au.

Film topography was analyzed in air with AFM (JPK NanoWizard®4 Bioscience), in QI-mode before and after the SFA experiments. A ContAl-G cantilever (NanoSensors, k = 0.2 N/m and l = 450 µm) was used to scan the surfaces (scan sizes of 2x2 and 15x15 µm$^2$. Both for SEM and AFM, the samples used in the SFA were quickly dried with $N_2$ after the experiments. The samples observed after the SFA experiments appeared cracked, but the cracking was caused by fast sample drying in a laminar flow cabinet (we would also observe such large cracks in the SFA camera if they appeared during the experiments).

## SFA Measurements and Data Analysis

Nm-range forces between two rough and polycrystalline calcite surfaces were measured with the SFA (SFA2000; SurForce LLC, USA [43]) as a function of a distance between the surfaces. Our SFA is coupled with MBI (Princeton Instruments IsoPlane SCT320 spectrometer and a PIXIS2048B camera with a lateral resolution of 0.62 μm/pixel), and a Thorlabs DCC1645C camera (0.15 μm/pixel resolution) aiding surface topography observation. The spectrometer was calibrated using an Hg lamp within a 520-630 nm spectral range and spectrometer gratings of three different resolutions (600, 1200, 1800 g/mm) were used, depending on the mica substrate thickness. MBI provides information about surface separation and topographic information *in situ* through the FECO fringes, which are sensitive to thickness and refractive index of the sample[46]. Calcite surfaces on mica substrate were glued to cylindrical glass disks with the radius of curvature R = 2 cm, which yielded nominal contact areas of 100-150 μm in diameter. The bottom surface was mounted on a force measuring spring, with a spring constant k = 2000 N/m. The principles of the SFA and MBI techniques have been described in[43,45,46,74]. For each experiment we used two fresh pieces of the ALD-deposited calcite films. We first established a suitable contact area, without visible, larger surface asperities, estimated the thickness of calcite surfaces, and then measured forces in the same contact throughout the 2-days (set 1) or 1-day (set 2) experiments. We analysed the SFA data using the open source Reflcalc software[75], which can simulate the FECO fringe patters by calculating the light transmission through our multi-layered samples. Identification of FECO wavelength positions and data processing has been handled in the MATLAB software. The details of Reflcalc modelling, data analysis, and typical experimental steps have been outlined in Dziadkowiec, et al. [21] and the **Supplementary Information** therein. We expect a relatively small error in determination of absolute separation between the surfaces for experiments in which we observed flattening of FECO fringes in contact (<20 nm, set 2 before PFs), and larger errors for rougher surfaces where the contact position was not reached in the range of applied loads that we used (even ~ 100 nm)[21]. The relative error between the consecutive data points in force curves, due to misestimation of absolute separation, should be less than several nm[21].

## Atomic Force Microscopy (AFM) measurements

Roughness evolution with time of single, unconfined calcite films in salt solutions was analyzed with the Atomic Force Microscope (AFM; MFP3D, Asylum Research, Oxford Instrument). A soft, uncoated quartz-like AFM tip with k = 0.01 N/m (qp-SCONT; NANOSENSORS™ uniqprobes) was used to image the surfaces in a contact mode (scan size 3x3 µm², resolution of 512 pixels). The experiments were carried out in stationary salt solutions, in a homemade, non-sealed fluid cell with a volume of ~ 3 ml. We thus observed some evaporation during the experiments, leading to an increase in salt concentration throughout the experiment. In each experiment we continuously scanned the same position on the film surface, however due to instrumental drift we usually observed a µm-range shift from the initial scan position. A new piece of calcite film deposited on mica (ALD set 3) was used for each experiment.

## Solutions

We used NaCl, $CaCl_2$ and $MgCl_2$ salt solutions with ionic strength of 10, 100 and 1000 mM. All solutions were presaturated with calcite by adding ~1 g/L of synthetic calcite powder (Merck KGaA; baked at 300˚C for 2 hours before use to reduce possible organic contamination). The salt/$CaCO_3$ solutions were sealed and stirred for more than one week prior to use. Prior to the SFA and AFM experiments, all solutions were filtered with 0.2 µm polyether-sulfone filters and injected into the SFA (~150 ml) or AFM (~3ml) directly after filtration. In the SFA the solutions were injected when keeping the two calcite surfaces in contact to limit dissolution upon equilibration with the solution. Every time when a new solution was injected into the SFA, the SFA chamber was drained with an excess ~150 ml of the same solution to limit possible contamination. The saturation indices (SI) with respect to calcite and $Ca^{2+}$ concentration were calculated in the PhreeqC software, using the $'llnl.dat'$ database, based on the measured pH and assuming $pCO_2$ both for closed ($10^{-6.2}$ atm) and open systems ($10^{-3.5}$ atm).

# Acknowledgements

This project has received funding from the European Union Horizon 2020 research and innovation program under the Marie Skłodowska-Curie grant agreement no. 642976-Nano-Heal Project. This work reflects only the author's view and the Commission is not responsible for any use that may be made of the information it contains. We thank Jon Einar Bratvold and Ola Nilsen for preparing ALD calcite films and providing XRD data. We acknowledge Shaghayegh Javadi and Agnès Piednoir for the useful advice and the help with AFM.

# Competing interests

The authors declare no competing interests.

# Data availability

The datasets generated during the current study are available from the corresponding author on reasonable request.

# Author contributions

JD wrote the manuscript, designed the experiments, performed SFA and AFM experiments, analyzed and interpreted the data; BZ performed AFM experiments and contributed to data analysis; DKD contributed to data interpretation and manuscript writing, AR supervised the project, designed the experiments, contributed to data analysis, interpretation, and manuscript writing.

# Supplementary Information for

# Nucleation in confinement generates long-range repulsive forces between rough calcite surfaces.

Joanna Dziadkowiec*,[1], Bahareh Zareeipolgardani[2], Dag Kristian Dysthe[1], Anja Røyne[1]

[1]Physics of Geological Processes (PGP), The NJORD Centre, Department of Physics, University of Oslo, Oslo 0371, Norway

[2]Institut Lumière Matière, Université de Lyon, Université Claude Bernard Lyon 1, CNRS UMR 5586, Campus de la Doua, F-69622 Villeurbanne cedex, France

* joanna.dziadkowiec@fys.uio.no

# Supplementary Movies

All supplementary movies can be downloaded from:

https://www.dropbox.com/sh/cd9svatoslenxh5/AABoKF-vtvCV_wJTZNwxFtqLa?dl=0

The field of view is 192 μm x 154 μm for the movies M1-M15. Magnification of the M16 movie is indicated with a scale bar.

Time is displayed with respect to injection times of electrolyte solutions into the SFA chamber.

**FR** – the movie was recorded during force measurements (continued loading-unloading cycles)

**TO** – the movie was recorded when the surfaces were kept in contact under constant load

**PF onset** – elapsed time after the solution injection when the precipitation front (PF) entered the observed region

## Movies showing precipitation fronts: Set 1 experiments

Movie 1: 0.01 M NaCl experiment ('M1_NaCl001M_set1.avi')

*PF onset: 9 h; TO.*

*Note a depletion of the precipitate in the observed region after 15h, and a second PF event after 17 h.*

Movie 2: 0.1 M NaCl experiment ('M2_NaCl01M_set1.avi')

*PF onset: 3 h 30 min; FR.*

Movie 3: 1 M NaCl experiment ('M3_NaCl1M_set1.avi')

*PF onset: 4 h 20 min; FR.*

Movie 4: 0.01 M $CaCl_2$ experiment ('M4_CaCl001M_set1.avi')

*PF onset: 2 h 18 min; FR.*

*The PF event was captured in the observed region only at its initial stage.*

Movie 5: 0.1 M $CaCl_2$ experiment ('M5_cacl01M_set1.avi')

*PF onset: 2 h 20 min; FR.*

Movie 6: 1 M $CaCl_2$ experiment ('M6_cacl1M_set1.avi')

*PF onset: 1 h 40 min; FR.*

Movie 7: 0.1 M $MgCl_2$ experiment ('M7_mgcl01M_set1.avi')

*PF onset: 12 h; TO.*

Movie 8: 1 M MgCl$_2$ experiment ('M8_mgcl1M_set1.avi')

*PF onset: 14 h 30 min; TO.*

## Movies showing precipitation fronts: Set 2 experiments

Movie 9: 0.01 M NaCl experiment ('M9_NaCl001_set2.avi')

*PF onset: 1 h 40 min; FR.*

Movie 10: 0.1 M NaCl experiment ('M10_NaCl01_set2.avi')

*PF onset: 1 h 20 min; FR.*

Movie 11: 1 M NaCl experiment ('M11_NaCl1_set2.avi')

*PF onset: 50 min; FR.*

*Precipitate forms discontinuous domains and is mobile on repeated loading-unloading cycles.*

Movie 12: 0.01 M CaCl$_2$ experiment ('M12_CaCl001_set2.avi')

*PF onset: 48 min; FR.*

Movie 13: 0.1 M CaCl$_2$ experiment ('M13_CaCl01_set2.avi)

*PF onset: 2h 10 min; FR.*

Movie 14: 1 M CaCl$_2$ experiment ('M14_CaC1_set2.avi)

*PF onset: 2h 40 min; FR.*

## Movie showing a full loading-unloading cycle after the PF

Movie 15: 0.1 M NaCl experiment, set 1 ('M15_run_afterPF.avi')

*SFA force measurement 26 h after the solution injection and 22 h 30 min after the PF.*

## Movie showing smashing of the precipitate

Movie 16: precipitate smashing ('M16_precipitate_smashing.avi')

*Surfaces were approached at very high loads using the manual SFA micrometer control. After several loading-unloading cycles, µm-sized spherical particles started to appear in the contact region. It was very likely that these particles represented CaCO$_3$ crystals.*

# Atomic Layer Deposition (ALD) Parameters

**Table S1.** *ALD deposition parameters for the set 1, set 2 and set 3 calcite surfaces grown on mica substrates using the F-120 Sat reactor from ASM Microchemistry by the procedure adapted from [Nilsen, et al. [1]](#) . The $Ca^{2+}$ organic precursor was $Ca(thd)_2$ (Volatec; 97%; Hthds = 2,2,6,6-tetramethylheptan-3,5-dione).*

| T (°C) | aimed thickness (nm) | sublimation T (°C) | deposition cycles ||||||| number of cycles |
|---|---|---|---|---|---|---|---|---|---|
| | | | $Ca(thd)_2$ pulse (s) | $N_2$ purge (s) | $O_3$ pulse (s) | $N_2$ purge (s) | $CO_2$ pulse (s) | $N_2$ purge (s) | |
| 300 | 100 | 195 | 3 | 2 | 3 | 2 | 3 | 2 | 2000 |

# X-ray Diffraction (XRD)

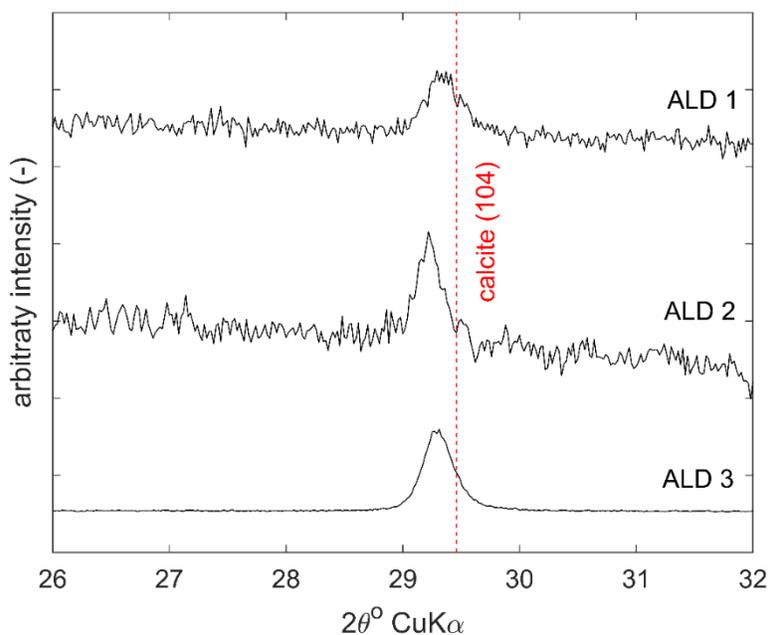

**Figure S1**. *XRD of ALD-deposited calcite films on Au substrate (ALD 1 - set 1, ALD 2 – set 2) and on Si wafer substrate (ALD 3 used in AFM; Figures S6, S7). Only the most intense calcite (104) peak can be identified for the films due to their small thickness (~100 nm). The position of the calcite peak is slightly shifted in the ALD films, which can be due to imperfect sample alignment (Au or Si substrates were attached to standard XRD holders). Other structural effects are also possible, however, because of very low peak intensity and small film thickness, such analysis is not feasible.*

# Scanning Electron Microscopy (SEM)

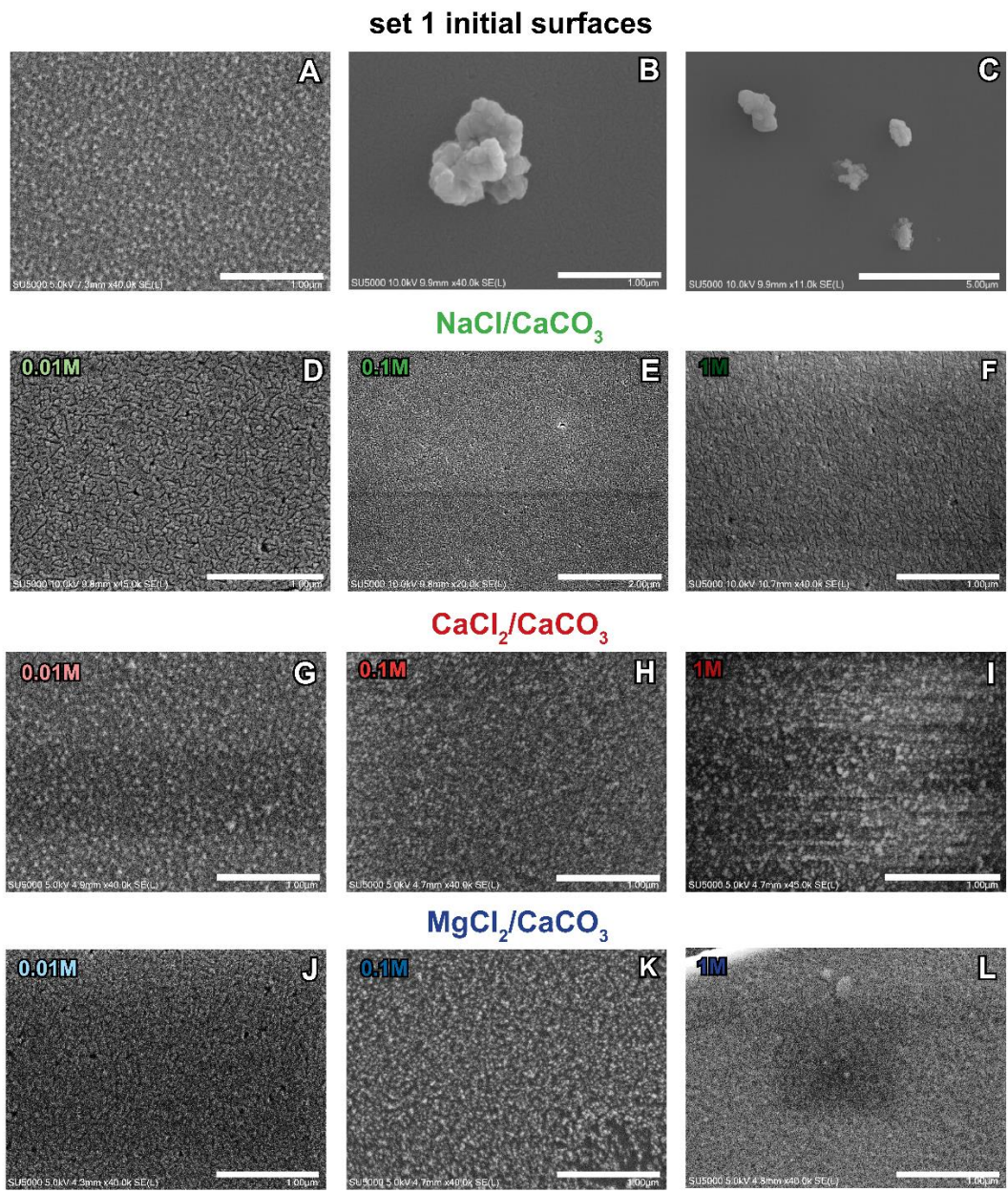

**Figure S2.** *SEM SE images of the initial (A-C) and final morphology (D-L) of the set 1 ALD calcite films. The D-L images show samples after the SFA experiments in NaCl, CaCl$_2$ and MgCl$_2$ solutions with ionic strength ranging from 0.01 to 1 M. Before the observations, samples were dried with N$_2$. Scale bars are 1 µm (C: 5 µm, E: 2 µm). Samples coated with Au. Visible CaCl$_2$ or MgCl$_2$ salt residue is visible on images H, I and K due to drying the wet samples after the SFA experiments with pressurized N$_2$.*
5

**set 2 initial surfaces**

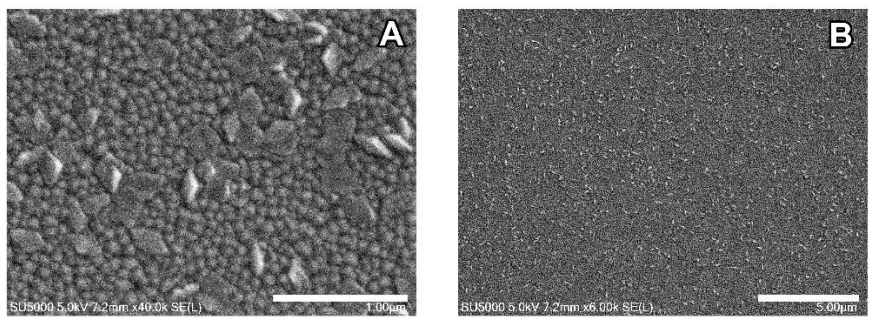

**NaCl/CaCO₃**

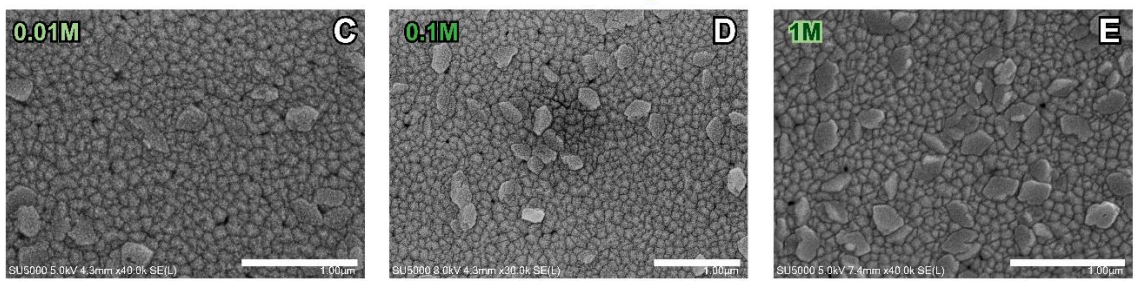

**CaCl₂/CaCO₃**

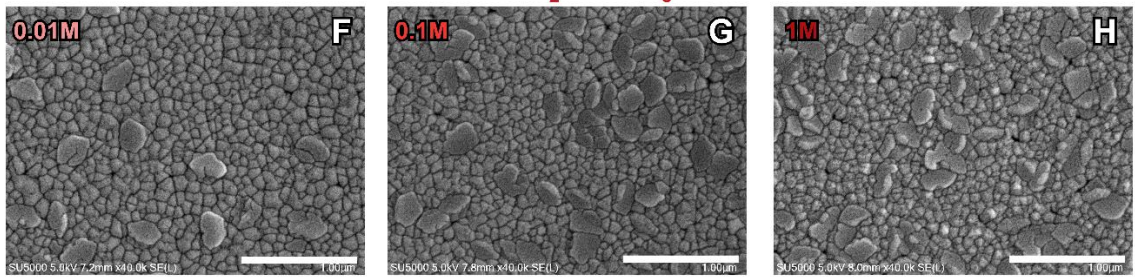

**MgCl₂/CaCO₃**

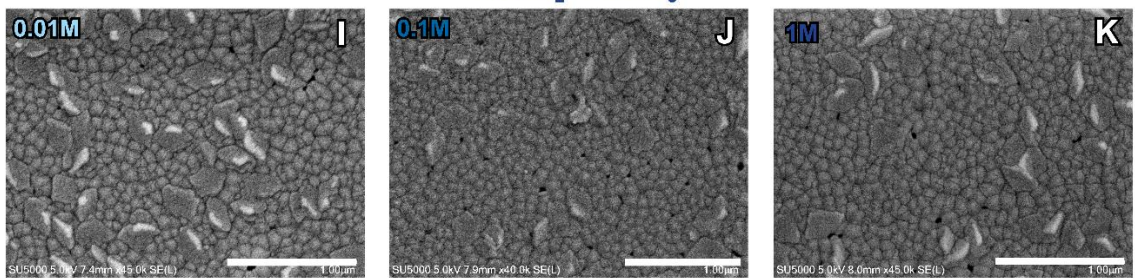

**Figure S3.** *SEM SE images of the initial (A-B) and final morphology (C-K) of the set 2 ALD calcite films. The C-K images show samples after the SFA experiments in NaCl, CaCl$_2$ and MgCl$_2$ solutions with ionic strength ranging from 0.01 to 1 M. Before the observations, samples were dried with N$_2$. Scale bars are 1 µm (B: 5 µm). Samples coated with Au.*

# Atomic Force Microscopy (AFM)
## ALD films roughness characterization

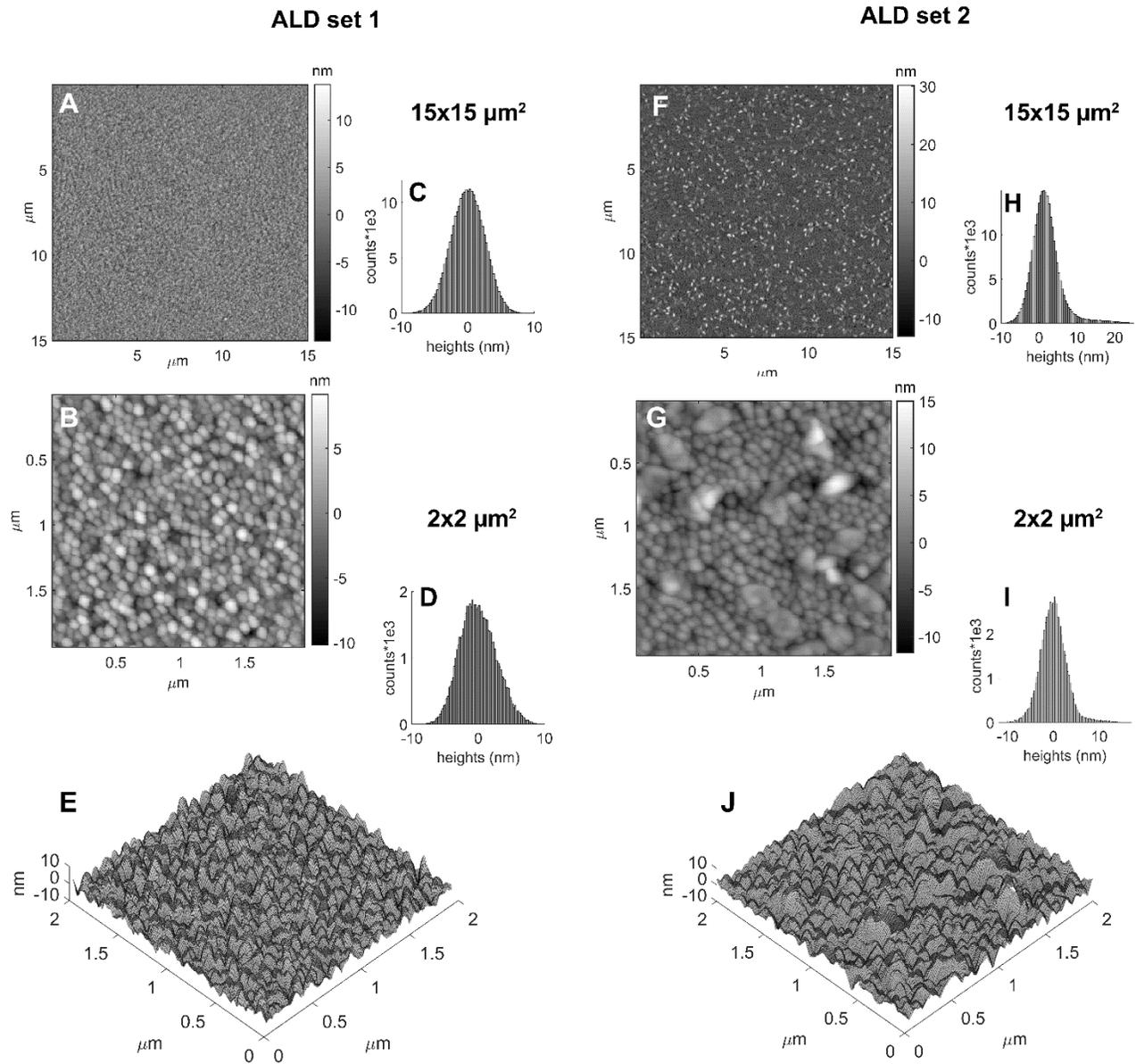

**Figure S4.** *AFM height maps (A, B, E, F, G, J) and histograms of surface heights (C, D, H, I) of the initial set 1 (A-E) and set 2 (F-J) ALD calcite surfaces for two scan sizes of 15x15 µm$^2$ (A, C, F, H) and 2x2 µm$^2$ (B, D, E, G, I, J). The images E and J show 3D height maps of the B, G height maps, respectively.*

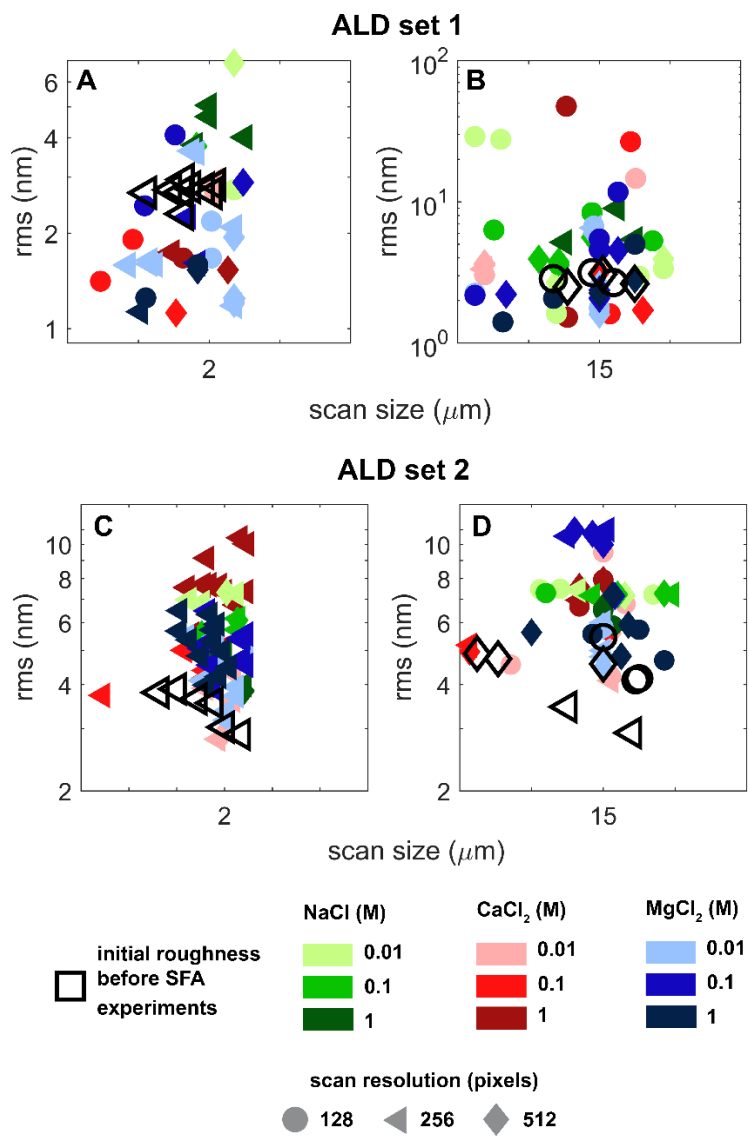

**Figure S5**. *AFM rms roughness parameters for the set 1 (A, B) and set 2 (C, D) ALD calcite surfaces for scan sizes of 15x15 µm² (B, D) and 2x2 µm² (A, C). Empty symbols correspond to rms values measured for the initial ALD surfaces before the SFA experiments. Colored symbols correspond to rms values measured for samples used in the SFA experiments in salt solutions with ionic strength (IS) varying from 0.01 to 1 M. For each sample we measured roughness in three random locations on the calcite surface, as it was not possible to locate where were the contacts used in the SFA measurements. Note a different y-scale in the B image. All salt solutions were presaturated with $CaCO_3$ as described in the Methods section.*

# Equilibration of single calcite surfaces with solutions in the AFM

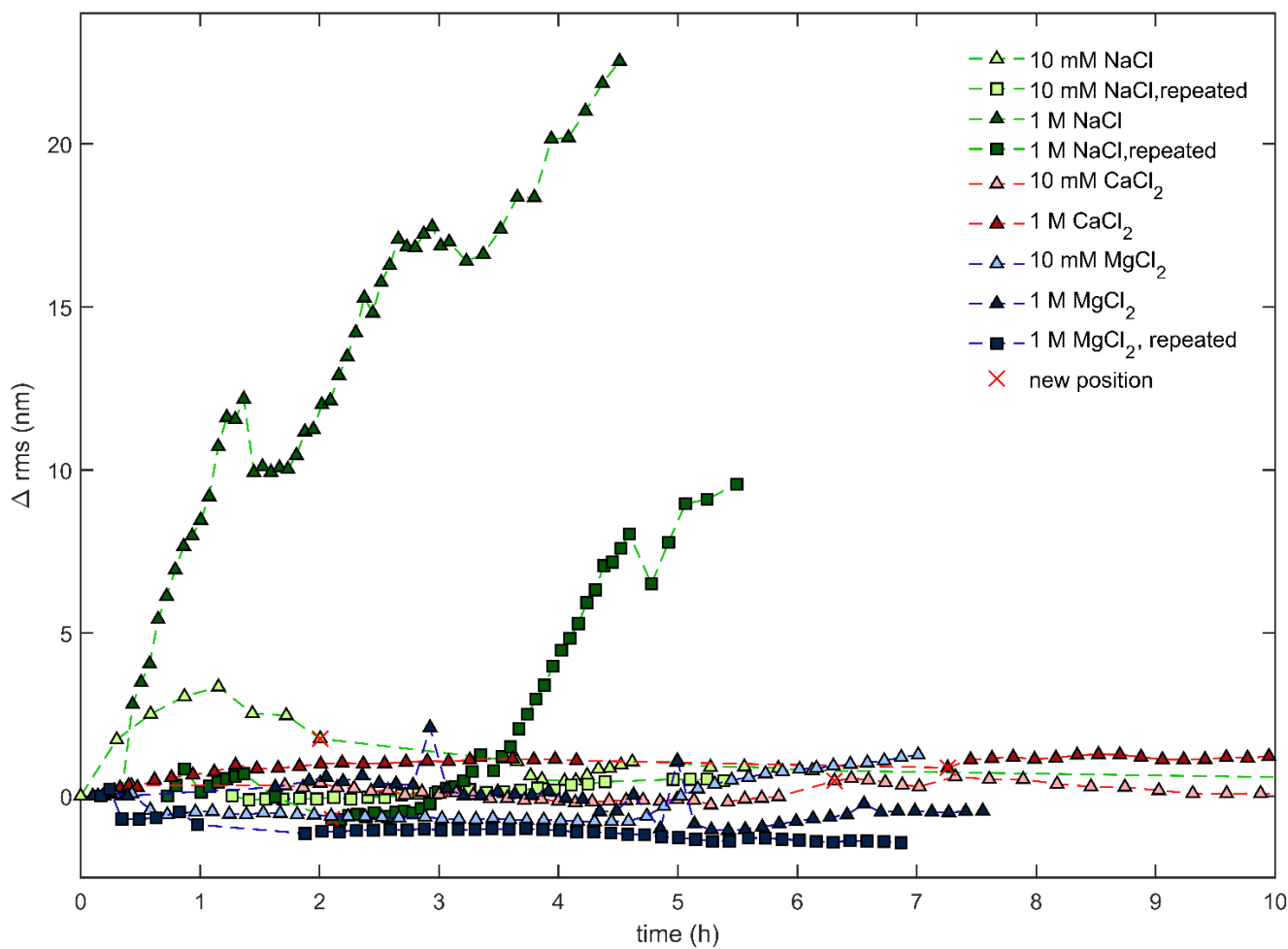

**Figure S6.** *AFM rms roughness evolution for the single, unconfined ALD calcite surfaces (set 3; scan size of 3x3 µm$^2$) in NaCl, CaCl$_2$ and MgCl$_2$ salt solutions with ionic strengths of 0.01 or 1 M. All salt solutions were presaturated with CaCO$_3$ as described in the Methods section. The red x symbols mark changes in scanning position on a sample whenever the signal was lost due to a large instrumental drift. We observed major changes in surface roughness (see Figure S7) only for the experiments in 1 M IS NaCl/CaCO$_3$ solutions. The details of the measurements are given in the Methods section.*

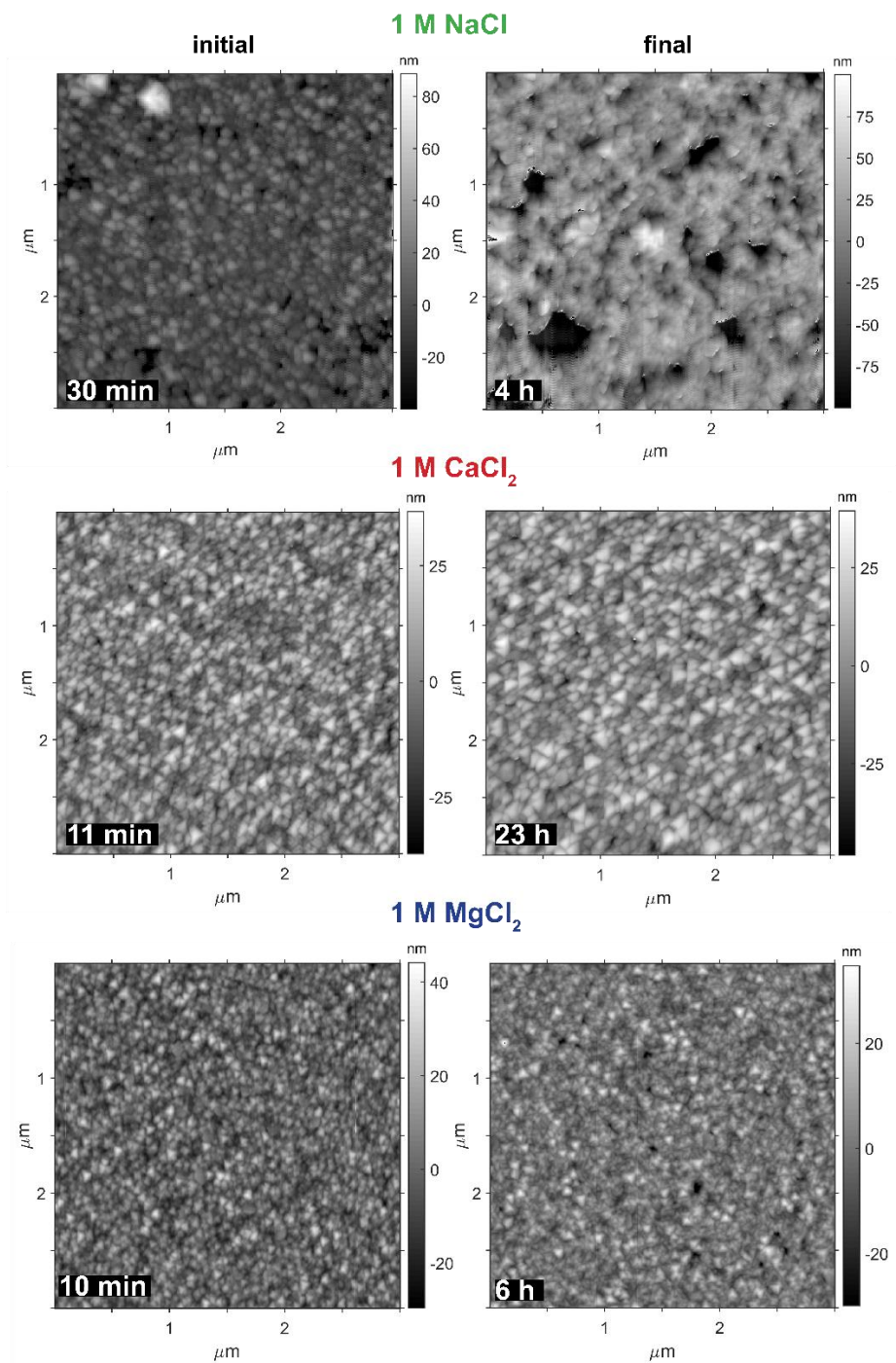

**Figure S7.** *AFM height maps for the set 3 ALD calcite surfaces (scan size of 3x3 µm²) at different stages of the AFM roughness evolution measurements with single unconfined calcite surfaces (corresponding to Figure S6). We observed major changes in surface morphology only for the most concentrated 1 M NaCl/CaCO$_3$ solutions, for which µm-sized dissolution pits were developing everywhere on the surface. We observed no major changes in topography for experiments in other 0.01 and 1 M ionic strength solutions.*

# Electrical Double Layer (EDL) Forces Modelling

EDL forces were estimated for calcite surfaces using the three following equations, assuming calcite surface potential ($\psi_0$) of 5 or 30 mV (Figure 8).

a)  Linear superposition approximation (LSA) method at constant potential (CP-LSA), adapted from Israelachvili [2] (see Figure 14.10, Chapter 14, page 317 therein):

$$EDL_{CP-LSA} = \kappa\sqrt{R^2}Ze^{-\kappa D}, \text{ where} \qquad \text{(Eq. S1)}$$

$$\kappa = \sqrt{\sum_i \frac{C_i e^2 z_i^2}{\varepsilon_0 \varepsilon kT}},$$

$$Z = 64\pi\varepsilon_0\varepsilon(kT/e)^2 tanh^2(\frac{ze\psi_0}{4kT}),$$

$\kappa^{-1}$ is Debye length (m$^{-1}$), $C_i$ is bulk concentration of each ion species $i$ in the solution (M), $z$ is ion valency, $\varepsilon_0$ is electrical permittivity of vacuum (F/m), $\varepsilon$ is the water dielectric constant, k is the Boltzmann constant, T is temperature (K), $R$ is the radius of the SFA cylindrical samples (m), and $D$ is the distance between the surfaces (m). For mixed 2:1 CaCl$_2$ and MgCl$_2$ we assumed $z$ = 2. For NaCl solutions we used $z$ = 1.

b)  Linearized Poisson-Boltzmann equation adapted from Diao and Espinosa-Marzal [3] and Trefalt, et al. [4] assuming a constant charge regulation parameter of calcite ($p_c$ = 0.62), that has been experimentally determined by Diao and Espinosa-Marzal [3] in a calcite-silica system:

$$EDL_{CR} = 4\pi R\varepsilon\varepsilon_0\kappa\psi_0^2 \frac{e^{-\kappa D}+e^{-2\kappa D}(2p_c-1)}{1-(2p_c-1)^2 e^{-2\kappa D}} \qquad \text{(Eq. S2)}$$

c)  simplified EDL force expression at low constant surface potential (<25 mV), suitable for mixed electrolytes, adapted from Israelachvili [2] (see Chapter 14, equation 14.56, page 318 therein):

$$EDL_{CP} = 4\pi R\varepsilon\varepsilon_0\kappa\psi_0^2 e^{-\kappa D} \qquad \text{(Eq. S3)}$$

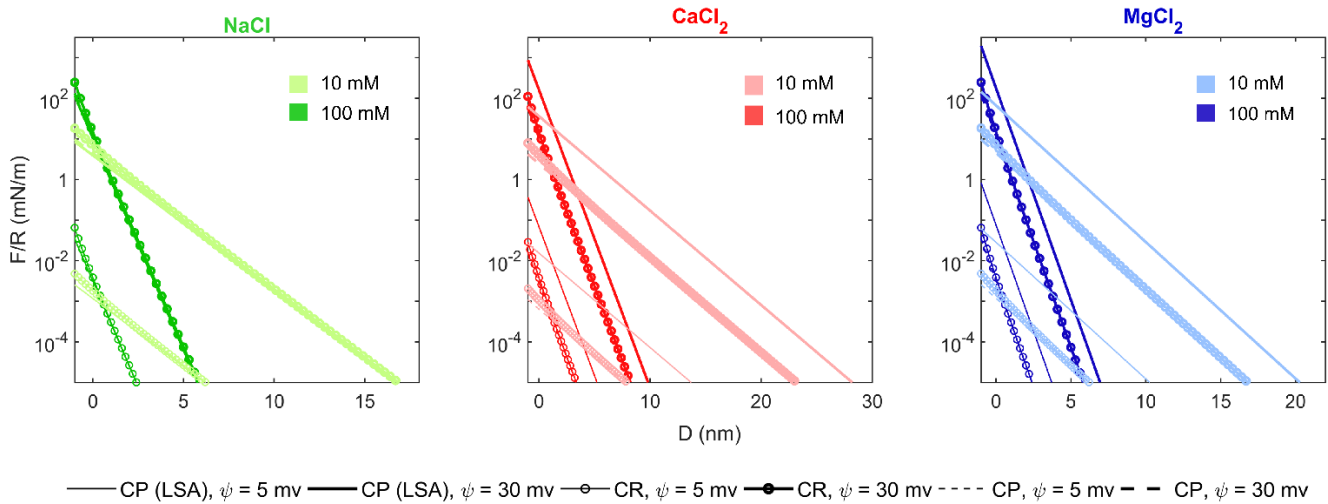

**Figure S8.** *Theoretical electrical double layer (EDL) force contributions estimated for two smooth calcite surfaces in 0.01 and 0.1 M IS electrolyte solutions. EDL forces were calculated using: a) linear superposition approximation (LSA) method at constant potential (CP (LSA); Eq. S1); b) linearized Poisson-Boltzmann equation at constant surface charge regulation using calcite surface charge regulation parameters adapted from [Diao and Espinosa-Marzal](#) [3] (CR, Eq. S2 ); c) simplified EDL force expression at constant surface potential, suitable for mixed electrolytes (CP, Eq. S3). EDL contributions were calculated assuming two values of calcite surface potential: 5 mV or 30 mV. Note that the EDL contributions calculated for NaCl solutions using all three expressions overlap, apart from the very small separations. Note that the EDL contributions calculated for $CaCl_2$ and $MgCl_2$ solutions using CR and CP expressions overlap, apart from the very small separations.*

# Surface Forces Apparatus (SFA) measurements

## Calcite Thickness

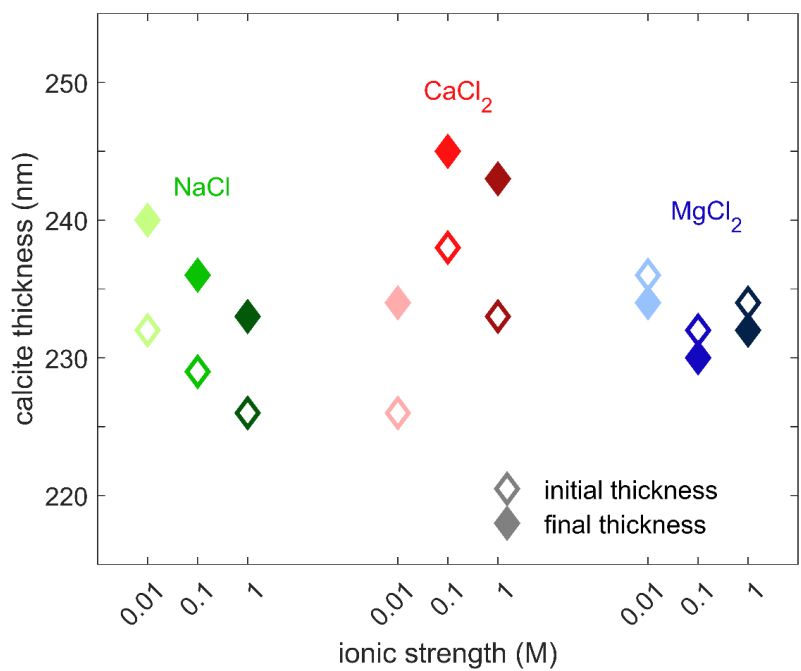

*Figure S9*. Calcite thickness values estimated for the set 2 calcite surfaces in contact regions chosen for the SFA force measurements. The initial calcite thickness (empty symbols) was measured by placing the two calcite surfaces in contact at small applied loads using the SFA motor-driven micrometer. These small loads were usually sufficient for the set 2 surfaces to flatten in contact. The initial flattening indicated that separation between the surfaces was nm-ranged over the whole nominal contact areas*. The final thickness (filled symbols) was estimated at the end of experiments. Because the precipitate was present between the surfaces, very high loads had to be applied in order to place the surfaces in flattened contacts. We used a manual SFA micrometer control to achieve these high loads. The very small difference in the initial and final calcite thicknesses for experiments in NaCl and $CaCl_2$ solutions indicates that most of the precipitate was squeezed out from between the surfaces at high applied loads. For the set 2 experiments in $MgCl_2$ solutions, for which we did not observe PFs, there was a decrease in calcite thickness in the contact region.

*Because of the nm-scale roughness of our ALD surfaces, only the highest asperities were in a direct contact, and separations varied across the large nominal contact areas (~100 μm in radius). Therefore, the plotted calcite thicknesses are only average values across the whole nominal contact areas, related to the distribution of the highest asperities. We also measured calcite thickness in one or two additional contact positions for each sample at the beginning of the experiments (not plotted here), and we obtained comparable calcite thickness values. That shows that the roughness and thickness of the set 2 calcite surfaces was homogenous over large areas of ALD films.

## Details of the SFA measurements

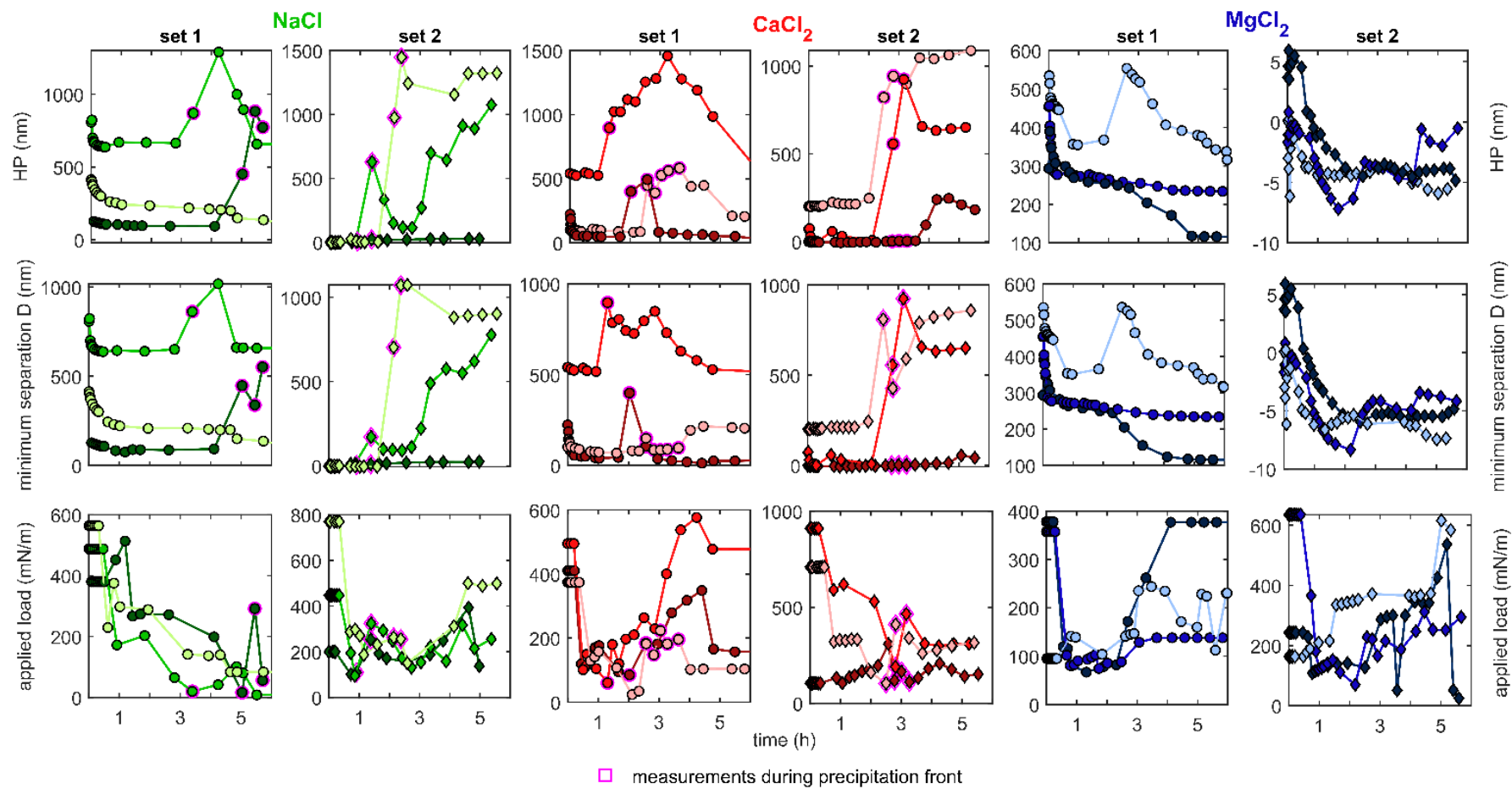

**Figure S10.** *Details of the SFA data shown in Figure 4. The middle panel shows minimum separation D at the maximum applied load as a function of elapsed time. The bottom panel shows maximum applied load as a function of elapsed time. Hardwall position (HP; top row) shows the separation between the surfaces at the applied load value common to all measurements for each experiment (the experimental points for each experiment are connected with solid lines; as in Figure 4). The colors correspond to ionic strength and composition of the used salt solutions that are consistent throughout the manuscript (e.g. Figure S5).*

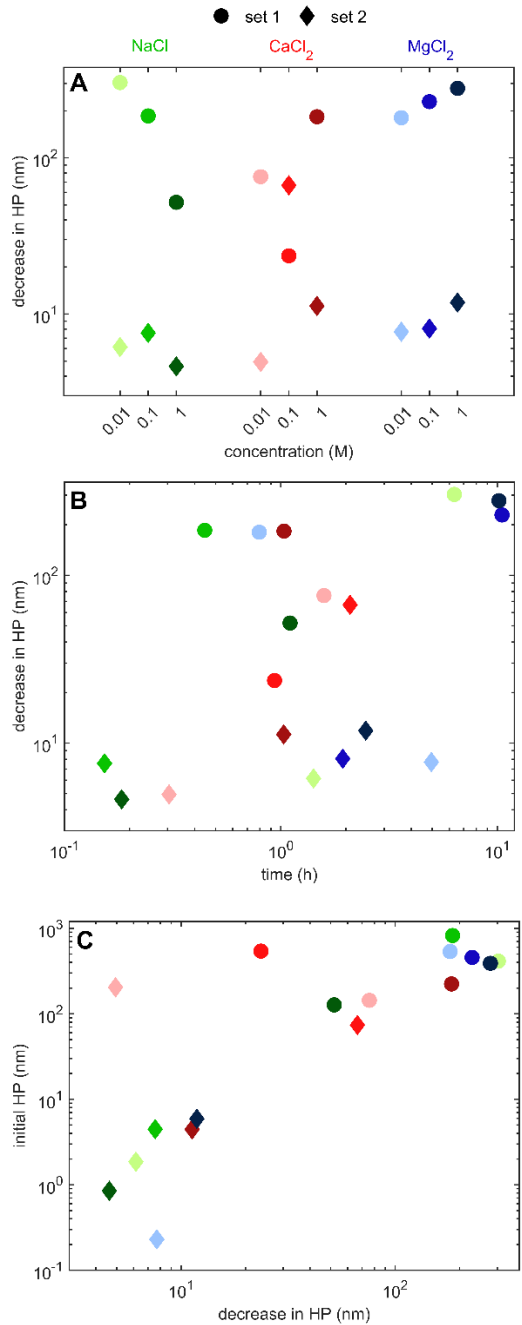

**Figure S11.** *Details of the SFA measurements for the set 1 (○) and set 2 (◊) surfaces (corresponding to the data shown in Figure 4). Colors indicate composition and ionic strength of the salt solutions used in the experiments (subplot A). A) Dependence of the decrease in HP measured before the PFs in the contact region used in the SFA experiments (or before the first progressive increase in HP for experiments in $MgCl_2$ solutions) on the ionic strength and composition of the used salt solutions; B) Dependence of the decrease in HP measured before the PFs in the contact region on the elapsed time; C) Dependence of the initial HP (first experimental point shown Figure 4 (top row) for each experiment on the decrease in HP measured before the PFs in the used contact region.*

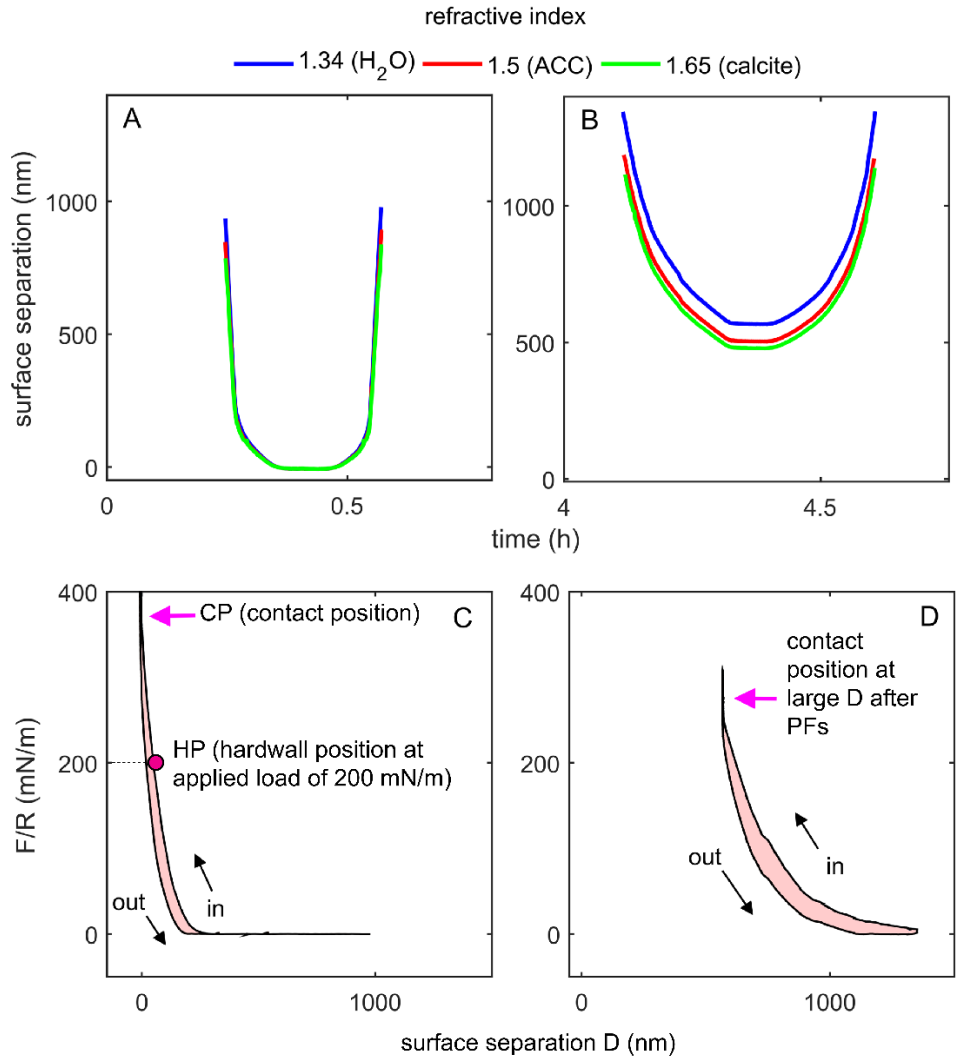

*Figure S12.* *Details of the SFA measurements showing effective surface separations modelled in Reflcalc (see the Methods section and [Dziadkowiec, et al. ⁵](#)) using different values of the refractive index ($n$) for the solution confined between two calcite surfaces. SFA data from the set 2 experiment in 0.1 M IS CaCl$_2$ solution (see Figure 4). A) Surface separation as a function of time for one representative loading-unloading cycle, measured before the PF event. At the beginning of the experiment it was possible to reach the initial CP. There was only a little difference between the surface separations calculated using the different $n$ values; B) Surface separation as a function of time for one representative loading-unloading cycle, measured after the PF event. At large surface separations, there was a substantial difference between the separations calculated using the three $n$ values. Even if the highest $n$ of calcite was used, the estimated surface separation was still large; C) Force-distance curve before PF, corresponding to the data in the subplot A. Locations of contact position (CP, separation at which distance between the surfaces no longer decreases despite continued loading) and hardwall position (HP, separation measured at a given applied load value) are indicated; D) Force-distance curve after PF, corresponding to the data in the subplot B. A contact position at very large separations is indicated, where the separation between the surfaces does not decrease further despite the continued loading.*

*Note that for water and calcite we used the tabulated values of n (adapted from Hale and Querry [6] and Ghosh [7], respectively). The values for birefringent calcite were used as an average value for ordinary and extraordinary rays at a given wavelength, whereas for ACC we used a constant value of 1.5 due to lack of detailed n parameters for ACC[8]. The legend shows values of n at wavelength of 600 nm for water and calcite (ordinary n).*

## SFA experiment in monoethylene glycol

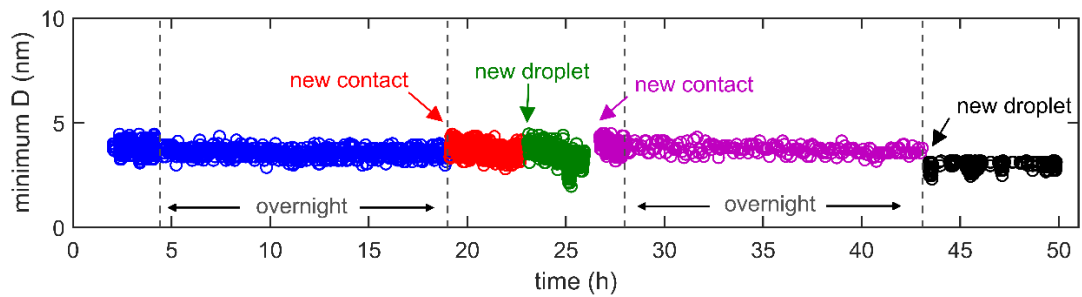

*Figure S13. Minimum separation between two calcite surfaces plotted as a function of elapsed time, measured in the SFA experiment in monoethylene glycol (MEG; ethane-1,2-diol; Merck, reagent grade, 99.5% pure). Data correspond to minimum separations measured during the consecutive force-distance runs or to periods of time when the surfaces were kept in contact under the constant applied load ('overnight'). Only a droplet of MEG (~ 2 ml) was injected between the surfaces. At the beginning of the experiment, MEG solution was exchanged multiple times to ensure a complete surface wetting. The arrows indicate when a contact position between the two calcite samples was changed or when MEG droplet was replaced with fresh MEG solution. MEG was not presaturated with calcite.*